\documentclass[journal]{IEEEtran}
\usepackage{graphicx,color}
\usepackage{amsmath,amssymb,amsthm}
\usepackage{cite}
\usepackage{times}
\usepackage{bm}
\usepackage{comment}
\usepackage{algorithmic,algorithm} 
\usepackage{url}
\usepackage{multirow}
\usepackage{arydshln}
\usepackage{mathtools}
\theoremstyle{definition}
\allowdisplaybreaks
\usepackage{colortbl}
\usepackage{lscape}
\usepackage{array}
\usepackage{flushend}

\usepackage{hyperref}
\usepackage{xcolor}
\usepackage{threeparttable}
\usepackage{makecell}
\usepackage{tabularx}

\usepackage{subfig}
\newcommand{\red}[1]{\textcolor{red}{#1}}

\newcommand{\Ccal}{\mathcal{C}}
\newcommand{\Ecal}{\mathcal{E}}

\newcommand{\Ncal}{\mathcal{N}}
\newcommand{\Pcal}{\mathcal{P}}
\newcommand{\Qcal}{\mathcal{Q}}
\newcommand{\Rcal}{\mathcal{R}}
\newcommand{\Scal}{\mathcal{S}}
\newcommand{\Vcal}{\mathcal{V}}

\newcommand{\relmiddle}[1]{\mathrel{}\middle#1\mathrel{}}

\newtheorem{problem}{Problem}
\newtheorem{remark}{Remark}

\begin{document}

\title{Policy-Driven Orchestration Framework \\ for Multi-Operator Non-Terrestrial Networks}

\author{
Yuma~Abe, Mariko~Sekiguchi, Go~Otsuru, and Amane~Miura       
\thanks{Y.~Abe, M.~Sekiguchi, G.~Otsuru, and A.~Miura are with the Space Communication Systems Laboratory, Wireless Networks Research Center, Network Research Institute, National Institute of Information and Communications Technology (NICT), Tokyo, Japan, e-mail: \{yuma.abe, sekiguchi, otsuru, amane\}@nict.go.jp}% <-this % stops a space
\thanks{Manuscript received xxx.}
}

% The paper headers
\markboth{Accepted for publication in IEEE Transactions on Communications}%
{Shell \MakeLowercase{\textit{et al.}}: Bare Demo of IEEEtran.cls for IEEE Journals}

% make the title area
\maketitle

\begin{abstract}
Non-terrestrial networks (NTNs) have gained significant attention for their scalability and wide coverage in next-generation communication systems.
A large number of NTN nodes, such as satellites, are required to establish a global NTN, but not all operators have the capability to deploy such a system.
Therefore, cooperation among multiple operators, facilitated by an orchestrator, enables the construction of virtually large-scale constellations.
In this paper, we propose a weak-control-based orchestration framework that coordinates multiple NTN operators while ensuring that operations align with the policies of both the orchestrator and the individual operators.
Unlike centralized orchestration frameworks, where the orchestrator determines the entire route from source to destination, the proposed framework allows each operator to select preferred routes from multiple candidates provided by the orchestrator.
To evaluate the effectiveness of our proposed framework, we conducted numerical simulations under various scenarios and network configurations including dynamic NTN environments with time-varying topologies, showing that inter-operator cooperation improves the availability of feasible end-to-end routes.
Furthermore, we analyzed the iterative negotiation process to address policy conflicts and quantitatively demonstrated the ``price of autonomy,'' where strict individual policies degrade global feasibility and performance.
The results also demonstrate that outcomes of the proposed framework depend on the operators' policies and that hop count and latency increase as the number of operators grows.
These findings validate the proposed framework's ability to deliver practical benefits of orchestrated multi-operator collaboration in future NTN environments.
\end{abstract}


\IEEEpeerreviewmaketitle

\section{Introduction}
\label{section:introduction}

\IEEEPARstart{I}{n} Beyond 5G/6G networks, non-terrestrial networks (NTNs) have gained significant attention because of their scalability and wide coverage~\cite{Al-Hraishawi_CST23, Abdelsadek_TC23, Nguyen_OJCS24}.
NTN consists of various communication platforms, hereafter referred to as NTN nodes, including geostationary orbit (GEO) satellites, non-geostationary orbit (NGSO) or low Earth orbit (LEO) satellites, high-altitude platform stations (HAPS), and drones.
This system interconnects diverse environments such as space, air, and maritime domains through various communication platforms.
Furthermore, by integrating NTN with terrestrial communication systems, users can achieve seamless connectivity regardless of their location.
Standardization efforts are currently underway within the Third Generation Partnership Project (3GPP), with specifications already being developed.

NTN networks and their traffic exhibit time-varying and heterogeneous characteristics.
To address these challenges, various network management methods and procedures have been proposed, e.g., a service-level agreement (SLA)-based route computation and resource allocation~\cite{Kak_TNSM22}, a slice-aware architecture and end-to-end slicing for seamless 5G-NTN integration~\cite{Drif_LCN21}, and artificial intelligence (AI) and machine learning (ML) techniques~\cite{Fontanesi_CST25, Bakambekova_OJCS24}.

\subsection{Motivation and Problem Context of NTN Orchestration}

\begin{figure}[tb]
	\centering
	\includegraphics[width=0.9\columnwidth]{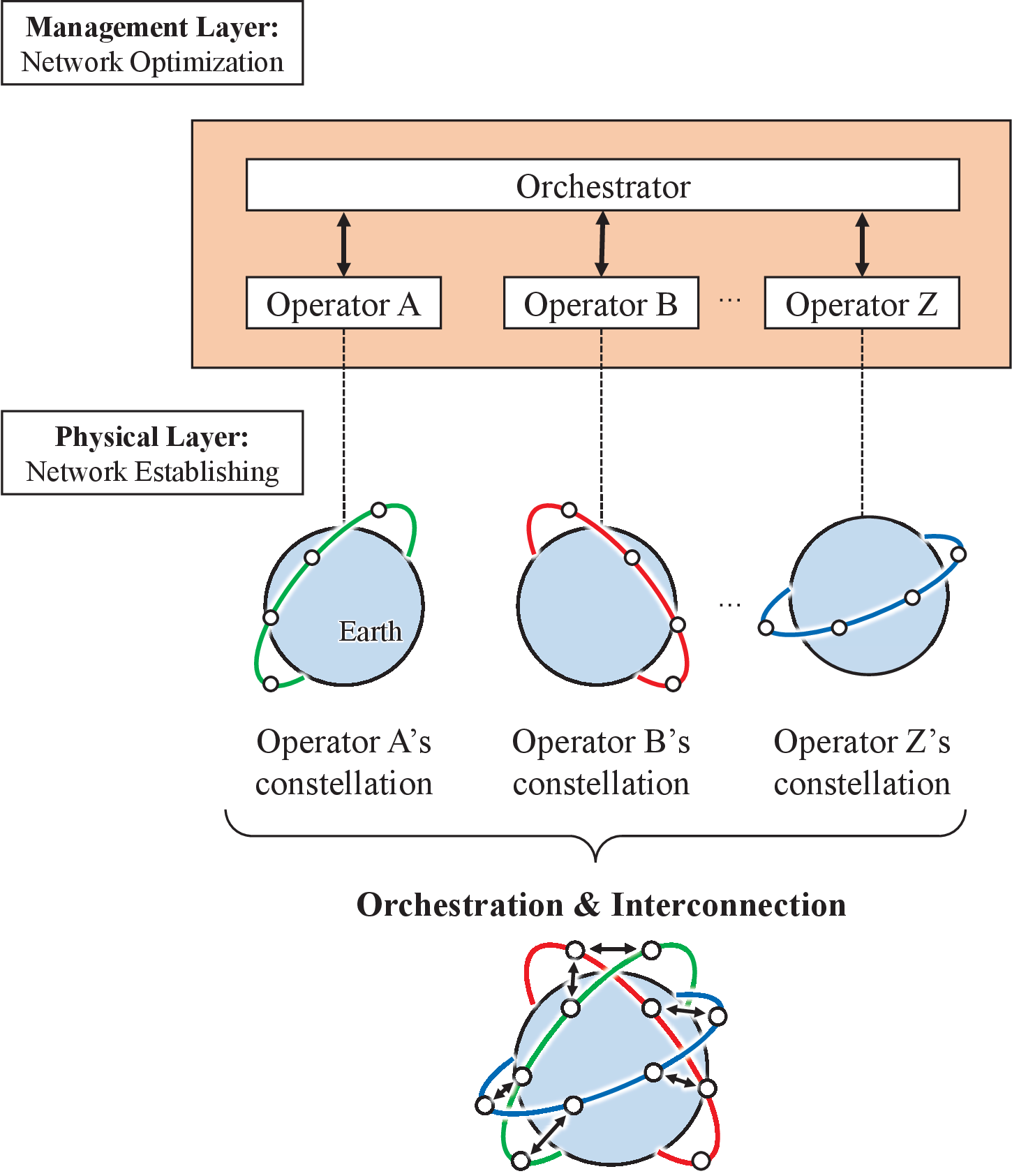}
	\caption{Conceptual diagram of orchestrating several constellations with a small number of satellites.}
\label{fig:constellation_orchestration}
\end{figure}

The rapid growth of NTN services is expected to involve a large number of users and numerous operators.
However, only a few operators can independently deploy large-scale constellations, such as SpaceX with its Starlink constellation exceeding 8,000 satellites as of October 2025~\cite{McDowell_Starlink}.
Most operators own only a few satellites and cannot provide full coverage independently.
This situation motivates cooperation among operators to construct virtual large-scale constellations through interconnection, as illustrated in Fig.~\ref{fig:constellation_orchestration}.
Such cooperation enables operators with limited satellite assets or those in the early stages of constellation deployment to provide services.

Direct coordination of inter-operator connections, i.e., links connecting nodes belonging to different operators, and connection points, i.e., nodes where traffic is handed over between operators, imposes a significant operational burden.
To address this challenge, an \textit{orchestrator} is introduced to coordinate multiple operators, facilitate network management, and enable efficient end-to-end route construction while reducing the operational burden.
Orchestration functions can generally be deployed either as a third-party entity or as part of an operator's internal management system; in this paper, we consider the former configuration.

The concept of orchestration has been widely explored in both terrestrial networks and NTN. 
It has also been incorporated into standardization efforts, such as the Integrated Network Control Architecture (INCA) in ITU-T~\cite{Kafle_CSM24, ITU-T_Y3207}, Management and Orchestration (MANO) in ETSI within the Network Function Virtualization (NFV) architecture~\cite{Hawilo_JSAC19}, and Service Management and Orchestration (SMO) in O-RAN~\cite{D'Oro_TMC24, Barritt_Aero22, Campana_EuCNC23, O-RAN-NTN_25}.
As an industrial example of this approach, Aalyria's Spacetime platform has recently been introduced to provide multi-orbit NTN orchestration~\cite{Uyeda_SIGCOMM22, Mowla_MECON-D2.1_25}. 
In addition, the European Union has launched the IRIS² (Infrastructure for Resilience, Interconnectivity and Security by Satellite) program, which also envisions multi-operator orchestration and service integration~\cite{EU_IRIS2, EUSPA-SATCOM23, Ferragut_IAC24}.

\subsection{Incorporating Operator Policies into NTN Orchestration}

In practice, NTN operators have heterogeneous capabilities, e.g., number of satellites, orbital configurations, and radio resources, and provide services under diverse SLAs.
They may also impose additional constraints during operation, such as avoiding interference or protecting their own assets~\cite{Fontanesi_CST25, Bakambekova_OJCS24}.
Moreover, operational policies often constitute confidential information, including proprietary know-how, and operators are generally unwilling to disclose them~\cite{He_TNSM22, Cai_TNSM24}.
Therefore, a practical orchestration framework must respect the privacy of operators while enabling their policies to be reflected in the end-to-end route construction.

These characteristics pose significant challenges for a conventional centralized orchestration framework.
An orchestrator is generally expected to guarantee that the minimum requirements for network operation are met.
However, if it centrally controls all domains and enforces its own policies, the resulting operations may deviate from the service policies of individual operators.
Moreover, such a framework assumes that operators disclose their policies to the orchestrator, which is often unrealistic. 
When an orchestrator is deployed as a third-party entity, it is also expected to ensure that decisions do not disproportionately favor any single operator.
Therefore, conventional centralized orchestration cannot adequately account for operator-specific policies, which motivates the development of a policy-driven framework.

To overcome these limitations, we propose a new orchestration framework based on the concept of ``weak-control''~\cite{Inoue_CSL19}. 
This concept was originally developed as a control framework for human-in-the-loop systems, where each decision-maker receives a set of permissible actions from a global controller and selects one according to its own preferences. 
This enables overall coordination without imposing a single centralized decision, while still allowing decision-makers to pursue their own objectives.
We leverage this idea in NTN orchestration by designing an orchestrator that offers feasible route options instead of enforcing one optimal route. 
Operators then select according to their own policies without disclosing them, and the orchestrator determines a ``suboptimal'' route that balances system-wide efficiency with operator autonomy.

\subsection{Main Contributions}

The contributions of this paper are summarized as follows:
\begin{itemize}
\item We propose a weak-control-based orchestration framework for coordinating multiple NTN operators and present its problem formulation along with the three-step procedure for solving it.
Unlike conventional centralized approaches, our framework allows operators to select routes according to their own policies without disclosing them, while the orchestrator ensures global feasibility and fairness.

\item We perform numerical simulations under diverse scenarios and network configurations including dynamic NTN environments with time-varying topologies, evaluating performance metrics such as end-to-end latency, hop count, the number of inter-operator connections, and the number of routes selected by the orchestrator and the operators.
Furthermore, we investigate the fundamental benefits of operator collaboration by analyzing the impact of varying the number of cooperating operators on overall network performance, as well as evaluating multi-layered scenarios that incorporate both GEO and LEO satellites.
Moreover, we analyze the iterative negotiation process to address policy conflicts and quantitatively demonstrate the ``price of autonomy,'' where strict individual policies degrade global feasibility and performance.
\end{itemize}

The remainder of this paper is organized as follows.
Section~\ref{section:system_model} introduces the system model used in this paper: an expression of operators, a set of nodes and links, and feasible link conditions.
Section~\ref{section:preliminaries} formulates the NTN orchestration problem under the conventional orchestration framework and presents an illustrative example to motivate the proposed approach.
Section~\ref{section:proposed_framework} presents the problem formulation of the proposed weak-control-based orchestration framework and the three-step procedure for solving it, together with illustrative examples as in Section~\ref{section:preliminaries}.
In Section~\ref{section:simulation}, we conduct five types of numerical simulations to evaluate the effectiveness of our proposed framework.
Section~\ref{section:conclusion} concludes the paper.

\smallskip
\textit{Notation}: $\mathbb{N}$, $\mathbb{R}^{+}$, and $\mathbb{R}^{n}$ denote the sets of natural numbers, positive real numbers, and $n$-dimensional real vectors, respectively.
For a link or a route in the network, the notations $p$, $\Pcal$, and $\bm{\Pcal}$ denote a link, a route, or a portion of a route (i.e., a set of links), and a set of routes, respectively.

\section{System Model}
\label{section:system_model}

We consider an NTN system with an orchestrator and multiple operators.
We define a set of operators as
\begin{align}
    \Ncal_{p} = \{1, 2, \ldots, N_{p} \}, \nonumber
\end{align}
where $N_{p}$ represents the number of operators.

We model the system as a graph $(\Vcal, \Ecal)$, where $\Vcal$ and $\Ecal$ represent the sets of all nodes and links, respectively. 
The set $\Vcal$ includes the source node $s$ and the destination node $d$ of the traffic, and $\Vcal_{i}$ denotes the set of nodes belonging to the $i$-th operator. 
We denote by $\Ecal_{i}$ the set of feasible links among nodes of the $i$-th operator and by $\Ecal_{(i,j)}$ the set of feasible links between nodes of the $i$-th and $j$-th operators. 
The feasibility of each link depends on factors such as relative satellite positioning, link status, weather conditions, and potential failures, and is determined by the required conditions described in the following paragraph. 
The entire sets of nodes and links are represented as $\Vcal = \bigcup_{i\in\Ncal_{p}} \Vcal_{i}$ and $\Ecal = \bigcup_{i\in\Ncal_{p}} \Ecal_{i} \cup \bigcup_{i\neq j} \Ecal_{(i,j)}$, respectively. 
The feasible link set is time-varying, but time-step subscripts are omitted here for notational simplicity.

We describe the link conditions and the method used to calculate feasible links.
The following conditions must be satisfied for a link to be considered feasible:
\begin{itemize}
    \item The transmitter of the sending node and the receiver of the receiving node are visible, and the distance between them is less than or equal to the distance threshold $D_{\text{req}}$.
    \item For radio frequency (RF) links, the carrier-to-noise ratio (CNR) at the receiving node must be greater than or equal to the required value, $C/N_{\text{req}}$.
    \item For optical links, the received power at the receiving node must be greater than or equal to the required power, $P_{r,\text{req}}$.
\end{itemize}

We describe the specific equation for the second and third conditions.
The received power $P_{r}$ by the receiver equals the transmitted power from the transmitter plus all gains minus all losses as expressed by~\cite{Elbert_08}
\begin{align}
    P_{r} = P_{t} + G_{t} - L_{f} + G_{r} - L_{s},
\label{eq:received_power}
\end{align}
where $P_{t}$ is the transmit power, $G_{t}$ is the transmit antenna gain, $L_{f}$ is the free-space loss, $G_{r}$ is the receive antenna gain, and $L_{s}$ is other losses that should be included such as atmospheric loss, rain attenuation, and pointing loss.
The unit of each parameter is decibels.
Note that in inter-satellite links, $L_{s}=0$.
The free space loss $L_{f}$ caused by the spatial propagation of signals attenuates the power of the signal by a factor of
\begin{align}
	L_{f} = \left(\frac{4\pi d}{\lambda}\right)^{2} = \left(\frac{4\pi df}{c}\right)^{2}, \nonumber
\end{align}
where $d$ is the distance between the transmitter and receiver, $\lambda$ is the wavelength, $f$ is the carrier frequency, and $c$ is the speed of light ($c=3.0\times10^{8}$~m/s).

To measure the quality of RF communication links, the CNR $C/N_{r}$, which represents the ratio of the received carrier power $P_{r}$ to the system noise power $N_{r}$, needs to be calculated.
The system noise of interest is mainly thermal noise, which acts as a disturbance that influences the received signal and degrades the quality of the signal in the receiver.
Thus, the received system noise power $N_{r}$ is represented by $N_{r} = k_{B} T_{s} B_{w}$, where $k_{B}$ is the Boltzmann constant, $k_{\text{B}} = 1.38 \times 10^{-23}$~J/K, $T_{s}$ and $B_{w}$ represent the system noise temperature and the carrier bandwidth, respectively.
Then, the CNR in RF links is denoted by $C/N = P_{r} - N_{r}$.
Therefore, the conditions for the RF links and optical links are described by $C/N \geq C/N_{\text{req}}$ and $P_{r} \geq P_{r,\text{req}}$, respectively.

The orchestrator collects information on each operator's NTN nodes, including the number of deployed communication terminals as well as their field of view and steering ranges. 
Based on $\Vcal_{i}$ and these node characteristics, the orchestrator computes the sets of feasible links $\Ecal_{i}$ and $\Ecal_{(i,j)}$ that satisfy the conditions described above. 
Alternatively, the intra-operator link sets $\Ecal_{i}$ may be computed by each operator and then provided to the orchestrator.

\section{Preliminaries: NTN Orchestration Problem}
\label{section:preliminaries}

In this section, we formulate the NTN orchestration problem, where the objective is to construct a network that delivers user-requested traffic from a source node to a destination node.
Note that the source or destination can be set as satellites, such as earth observation satellites, which work as source nodes that collect some information and try to send the information to the ground.
To this end, we introduce the decision variables, feasibility constraints, and the policies defined by each operator and by the orchestrator. 
We then present the conventional centralized formulation and provide an illustrative example that serves as a baseline for the proposed method in the next section.

\subsection{Decision Variables and Constraints}
\label{section:variables_constraints}

We define a binary variable of link assignment for traffic, $x_{e}$, as follows:
\begin{align}
    x_{e} =
    \begin{cases}
        1, & \text{if traffic is assigned to a link $e\in\Ecal$,} \\
        0, & \text{otherwise.}
    \end{cases}
\nonumber
\end{align}
Here, a route, a set of links assigned to traffic, from the source to the destination, is represented by $\Rcal = \{e \mid x_{e} = 1\}\subseteq \Ecal$.
Let $\ell$ denote the latency of the traffic, calculated using the defined $x_{e}$ as:
\begin{align}
    \ell = \sum_{e\in\Rcal} \bar{\ell}_{e} x_{e}, \nonumber
\end{align}
where $\bar{\ell}_{e}$ denote the propagation latency of link $e$.

We introduce binary indicator variables $y_{e}$ and $z_{v}$ to model inter-operator connections and node avoidance, respectively:
\begin{align}
    y_{e} &=
    \begin{cases}
        1, & \text{if traffic is assigned to a link $e\in\Ecal_{(i,j)}$,} \\
        0, & \text{otherwise.}
    \end{cases} \nonumber \\
    z_{v} &=
    \begin{cases}
        1, & \text{if a node $v$ is used in a route $\Rcal$,} \\
        0, & \text{otherwise.}
    \end{cases} \nonumber
\end{align}

We require that the following flow conservation constraint hold:
\begin{align}
    \sum_{e\in\delta^{+}(v)} x_{e} - \sum_{e\in\delta^{-}(v)} x_{e} =
    \begin{cases}
        1, & \text{if $v=s$,} \\
        -1, & \text{if $v=d$,} \\
        0, & \text{otherwise,}
    \end{cases}
    ~~\forall v\in\Vcal
\label{eq:constraint_flow_conservation}
\end{align}
where $\delta^{+}(v)$ and $\delta^{-}(v)$ represent sets of outgoing/incoming links from/to node $v$, respectively.
Moreover, we ensure that no routing loop is formed in any selected link, represented by
\begin{align}
    \sum_{\substack{e\in\delta^{+}(v)}} x_{e} \leq 1, \sum_{\substack{e\in\delta^{-}(v)}} x_{e} \leq 1,~~\forall v\in\Vcal\setminus\{s, d\}.
\label{eq:constraint_no_loop}
\end{align}

\subsection{Policy}
\label{section:policy}

As discussed in the introduction, the orchestrator and each operator have their own network operation policies.
To represent these policies in terms of performance indicators and constraints, we denote $J_{\pi}$ and $\Ccal_{\pi}$ as the evaluation function and the set of constraints for routes characterized by a policy $\pi$.
The policies of the orchestrator and each operator are denoted by $\pi_{0}$ and $\pi_{i}$, respectively.
For simplicity, we also define simplified expressions of the evaluation function, $J_{0}=J_{\pi_{0}}$ and $J_{i}=J_{\pi_{i}}$, and those of the constraints, $\Ccal_{0}=\Ccal_{\pi_{0}}$ and $\Ccal_{i}=\Ccal_{\pi_{i}}$.
Note that the set of constraints $\Ccal_{0}$ includes the fundamental constraints in Eqs.~(\ref{eq:constraint_flow_conservation})-(\ref{eq:constraint_no_loop}).

In this paper, the following are example policies that can be set by the orchestrator and the operators:
\begin{itemize}
    \item $\pi_{\phi}$: No policy; all route candidates are acceptable.
    \item $\pi_{\text{ML}}$: Prefers routes with minimized latency.
    \item $\pi_{\text{LL}}(n_{l})$: Prefers routes with latency less than or equal to $n_{l}\in\mathbb{R}^{+}$.
    \item $\pi_{\text{MH}}$: Prefers routes with the minimized hop count.
    \item $\pi_{\text{LH}}(n_{h})$: Prefers routes with the hop count less than or equal to $n_{h}\in\mathbb{N}$.
    \item $\pi_{\text{MO}}$: Prefers a route with the minimized number of inter-operator connections.
    \item $\pi_{\text{LO}}(n_{o})$: Prefers routes with the number of inter-operator connections less than or equal to $n_{o}\in\mathbb{N}$.
    \item $\pi_{\text{AN}}(\Vcal_{a})$: Prefers routes that avoid specific nodes denoted as $\Vcal_{a}\subseteq\Vcal$. Multiple nodes can also be specified.
\end{itemize}

\noindent
The mathematical expressions of the policies are listed in Table~\ref{table:policy_list}. 
Using the decision variables introduced in Section~\ref{section:variables_constraints}, the evaluation functions of $\pi_{\text{ML}}$, $\pi_{\text{MH}}$, and $\pi_{\text{MO}}$ are defined as 
$J_{\pi_{\text{ML}}} = \sum_{e\in\Rcal} \bar{\ell}_{e} x_{e}$, 
$J_{\pi_{\text{MH}}} = \sum_{e\in\Rcal} x_{e}$, 
$J_{\pi_{\text{MO}}} = \sum_{e\in\Rcal} y_{e}$, respectively, 
where $\gamma_{v}>0$ denotes the penalty weight for using node $v$.
The policies $\pi_{\text{LL}}(n_{l})$, $\pi_{\text{LH}}(n_{h})$, and $\pi_{\text{LO}}(n_{o})$ are expressed as constraints and are defined using the above evaluation functions as $J_{\pi_{\text{ML}}} \leq n_{l}$, $J_{\pi_{\text{MH}}} \leq n_{h}$, and $J_{\pi_{\text{MO}}} \leq n_{o}$, respectively.
The policy $\pi_{\text{AO}}(\Vcal_{a})$ can be expressed either by the evaluation function $J_{\pi_{\text{AO}}} = \sum_{v\in\Vcal_{a}} \gamma_{v} z_{v}$ or by the constraint $z_{v}=0 \; (v\in\Vcal_{a})$.
For notational convenience, we denote the upper bounds of the evaluation functions that serve as constraints for the orchestrator and the operators (namely $n_{l}$, $n_{h}$, and $n_{o}$ in the above examples) by $\bar{J}_{0}$ and $\bar{J}_{i}$, respectively.
Throughout this paper, we use the logical conjunction operator ``$\land$'' to denote the combination of multiple policies. 
For example, a combined policy $\pi = \pi_{1} \land \pi_{2}$ indicates that both policies $\pi_{1}$ and $\pi_{2}$ are applied simultaneously.

\begin{table}[tb]
    \centering
    \caption{Policies and their mathematical expressions.}
    \scalebox{1}{
        \begin{tabular}{lll} \hline
            Policy $\pi$ & Evaluation function $J_{\pi}$ & Constraint $\Ccal_{\pi}$ \\ \hline \hline
            $\pi_{\text{ML}}$ & $J_{\pi_{\text{ML}}} = \sum_{e\in\Rcal} \bar{\ell}_{e} x_{e}$ & - \\
            $\pi_{\text{LL}}(n_{l})$ & - & $J_{\pi_{\text{ML}}} \leq n_{l}$ \\
            $\pi_{\text{MH}}$ & $J_{\pi_{\text{MH}}} = \sum_{e\in\Rcal} x_{e}$ & - \\
            $\pi_{\text{LH}}(n_{h})$ & - & $J_{\pi_{\text{MH}}} \leq n_{h}$ \\
            $\pi_{\text{MO}}$ & $J_{\pi_{\text{MO}}} = \sum_{e\in\Rcal} y_{e}$ & - \\
            $\pi_{\text{LO}}(n_{o})$ & - & $J_{\pi_{\text{MO}}} \leq n_{o}$ \\
            $\pi_{\text{AN}}(\Vcal_{a})$ & $J_{\pi_{\text{AO}}} = \sum_{v\in\Vcal_{a}} \gamma_{v} z_{v}$ & $z_{v}=0 \; (v\in\Vcal_{a})$
            \\
            \hline
        \end{tabular}
    }
\label{table:policy_list}
\end{table}

\subsection{Problem Setting: Conventional Centralized NTN Orchestration}

As discussed in the introduction, conventional orchestration is typically based on a centralized control framework.
In this framework, the orchestrator collects the network information from all operators and determines the end-to-end route in a centralized manner.
Within the conventional orchestration framework, and using the decision variables and feasibility constraints defined in the previous subsection, we formulate the optimization problem as follows:
\begin{problem}
    Given $\{\Vcal_{i}, \Ecal_{i}, \Ecal_{(i,j)}\}, i\in\Ncal_{p},j\neq i$, find $\Rcal^{*}$ that minimizes $J_{\pi_{0}}(\Rcal)$ such that $\Rcal\in\Ccal_{0}$.
\label{problem:conventional_orchestration}
\end{problem}

By solving this problem, the orchestrator specifies the route $\Rcal^{*}$ and notifies each operator to follow the resulting decision. 
Each operator then establishes its own network according to the orchestrator's instruction. 
However, since the orchestrator's policy $\pi_{0}$ does not necessarily align with those of the individual operators $\pi_{i}$, the resulting solution may not satisfy the policies of each operator.
This limitation motivates the proposed framework presented in the next section.

\subsection{Illustrative Example of Centralized Orchestration: Two Operators Case}
\label{section:centralized_orchestration_example}

In this subsection, we provide an example of the centralized orchestration introduced in the previous section.
We consider a scenario with an orchestrator and two NTN operators that aim to determine the optimal end-to-end traffic route from ground station A (the source, $s$) to ground station B (the destination, $d$), including the inter-operator connections.
We call operators that provide the first and second networks through which traffic passes as Operators A and B, respectively.

We define examples of evaluation functions of the orchestrator and operators.
Let $\Rcal$ be a variable representing end-to-end routes between the operators, while let $\Pcal_{\text{A}}$ and $\Pcal_{\text{B}}$ be variables representing routes within the networks of Operators~A and B, respectively.
We set an orchestrator's policy for network establishment management as $\pi_{0}=\pi_{\text{ML}}$, and an evaluation function based on this policy is represented by
\begin{align}
    J_{0}(\Rcal) &= \sum_{e\in\Rcal} \bar{\ell}_{e} x_{e}.
\label{eq:cost_function_example_orchestrator}
\end{align}
On the other hand, we define the policies of each operator as different from the orchestrator, and those of Operators A and B are represented as $\pi_{\text{A}}=\pi_{\text{AN}}(\Vcal_{a,\text{A}})$ and $\pi_{\text{B}}=\pi_{\text{MH}}$, respectively.
Evaluation functions defined under these policies are represented by
\begin{subequations}
\label{eq:cost_function_example_operators} \\
\begin{align}
    J_{\text{A}}(\Pcal_{\text{A}}) &= \sum_{v\in\Vcal_{a,\text{A}}} \gamma_{v} z_{v}, 
    \label{eq:cost_function_example_operator_a} \\
    J_{\text{B}}(\Pcal_{\text{B}}) &= \sum_{e\in\Rcal} x_{e},
    \label{eq:cost_function_example_operator_b}
\end{align}
\end{subequations}
respectively, where $\Vcal_{a,\text{A}}$ denotes a set of nodes that Operator~A wants to avoid.

\begin{figure}[tb]
	\centering
	\includegraphics[width=0.97\columnwidth]{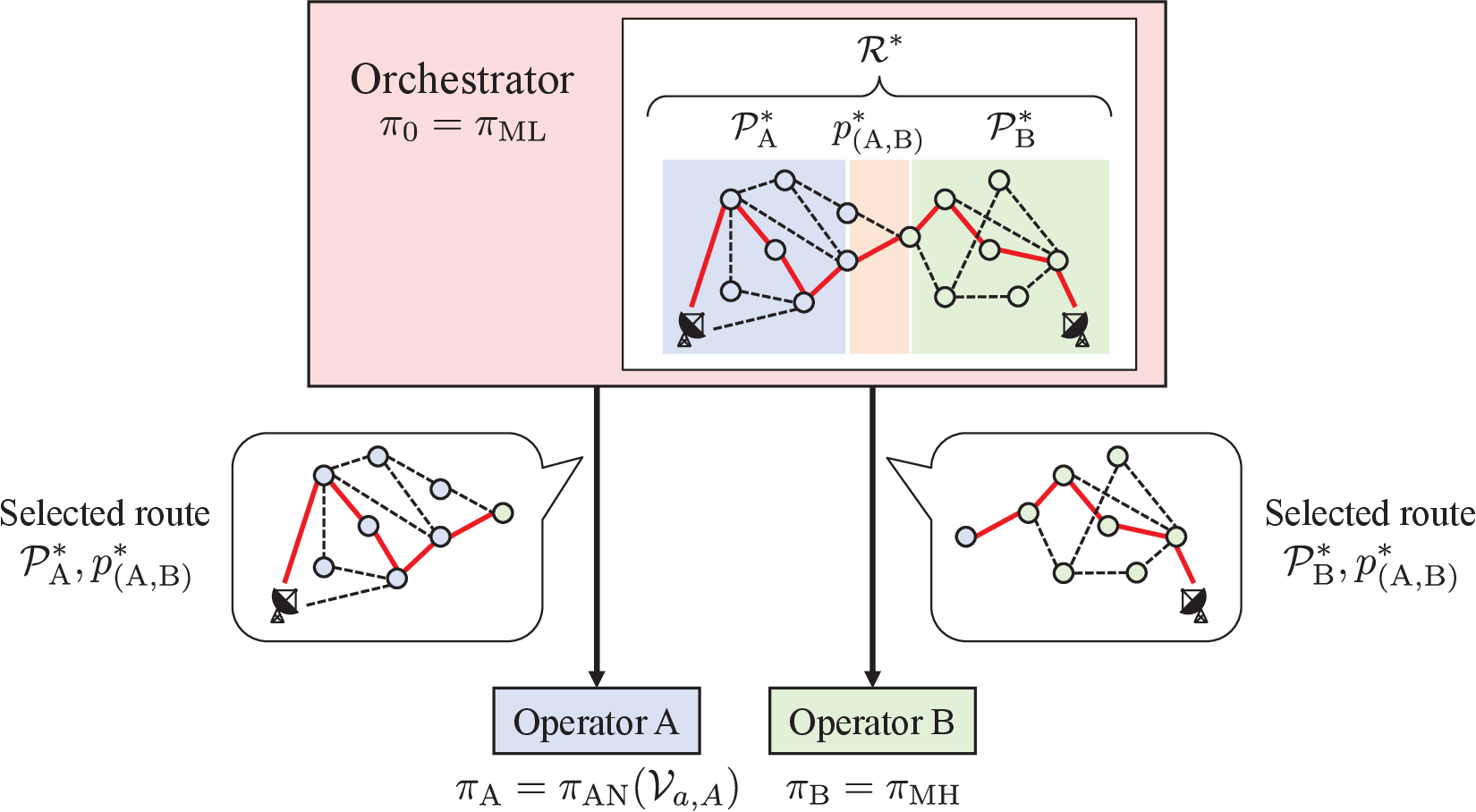}
	\caption{Conceptual diagram of the centralized orchestration framework in this example. Each circle represents NTN nodes, e.g., satellites.}
\label{fig:centralized_framework_example}
\end{figure}

The conceptual diagram of the centralized orchestration framework in this example is shown in Fig.~\ref{fig:centralized_framework_example}.
Based on the collected information on the NTN nodes $\Vcal_{i}$ owned by Operators A and B and the NTN links $\Ecal_{i}$ and $\Ecal_{(i,j)}$ formed by these nodes, the orchestrator solves Problem~\ref{problem:conventional_orchestration}:
\begin{align}
    \begin{aligned}
        & \underset{\substack{\Rcal}}{\text{minimize}}~~~
        && J_{0}~\text{in Eq.~(\ref{eq:cost_function_example_orchestrator})} \nonumber \\
        & \text{subject~to}
        && \Rcal \in \Ccal_{0}
    \end{aligned}
\end{align}
to obtain the optimal route $\Rcal^{*}$.
This route is divided as $\Rcal^{*} = \{\Pcal_{\text{A}}^{*}, p_{(\text{A},\text{B})}^{*},\Pcal_{\text{B}}^{*} \},$ where $\Pcal_{\text{A}}^{*}$ and $\Pcal_{\text{B}}^{*}$ represents the routes within the networks of Operators A and B, respectively, and $p_{(\text{A},\text{B})}^{*}$ represents the link between the two operators.
Then, the orchestrator notifies each operator of the selected route.
According to this notification, each operator follows the orchestrator's decision, establishes the network, and provides a communication link to the user.

In this case, the policy of the orchestrator is satisfied because the network is constructed based on the solution of the optimization problem concerning the orchestrator's evaluation function $J_{0}$.
On the other hand, the solution $(\Pcal_{\text{A}}^{*}, \Pcal_{\text{B}}^{*})$ does not always minimize $J_{\text{A}}(\Pcal_{\text{A}})$ in Eq.~(\ref{eq:cost_function_example_operator_a}) and $J_{\text{B}}(\Pcal_{\text{B}})$ in Eq.~(\ref{eq:cost_function_example_operator_b}) because the objectives of the orchestrator, Operator A, and Operator B are different, i.e., $J_{0} \neq J_{\text{A}}$ and $J_{0} \neq J_{\text{B}}$.
Therefore, the policies of Operators A and B may not have been satisfied.

\section{Proposed Orchestration Framework: Weak-Control-Based NTN Orchestration}
\label{section:proposed_framework}

This section formulates the proposed weak-control-based NTN orchestration framework.

As explained in the previous section, Problem~\ref{problem:conventional_orchestration} is insufficient because the policies of operators have been considered.
We then formulate the following problem, which considers policies of the operators:
\begin{problem}[Policy-Based Orchestration Problem]
    Given $\{\Vcal_{i}, \Ecal_{i}, \Ecal_{(i,j)}\}, i\in\Ncal_{p},j\neq i$, find $\Rcal^{*}$ that minimizes $J_{\pi_{0}}(\Rcal)$ such that $\Rcal\in\Ccal_{0}\cap\bigcap_{i\in\Ncal_{p}} \Ccal_{i}$.
\label{problem:weak_control_orchestration}
\end{problem}

However, this problem, which explicitly considers operator policies, cannot be solved directly in practice.
The policy of each operator $\pi_{i}$ is unknown to the orchestrator because operational policies may constitute confidential information, and thus operators are generally unwilling to disclose them~\cite{He_TNSM22, Cai_TNSM24}.
Therefore, the orchestrator cannot explicitly consider the operators' policies in the optimization problem.

\begin{figure}[tb]
    \centering
    \begin{minipage}[t]{0.98\hsize}
        \centering
        \subfloat[]{
            \includegraphics[width=0.95\columnwidth]{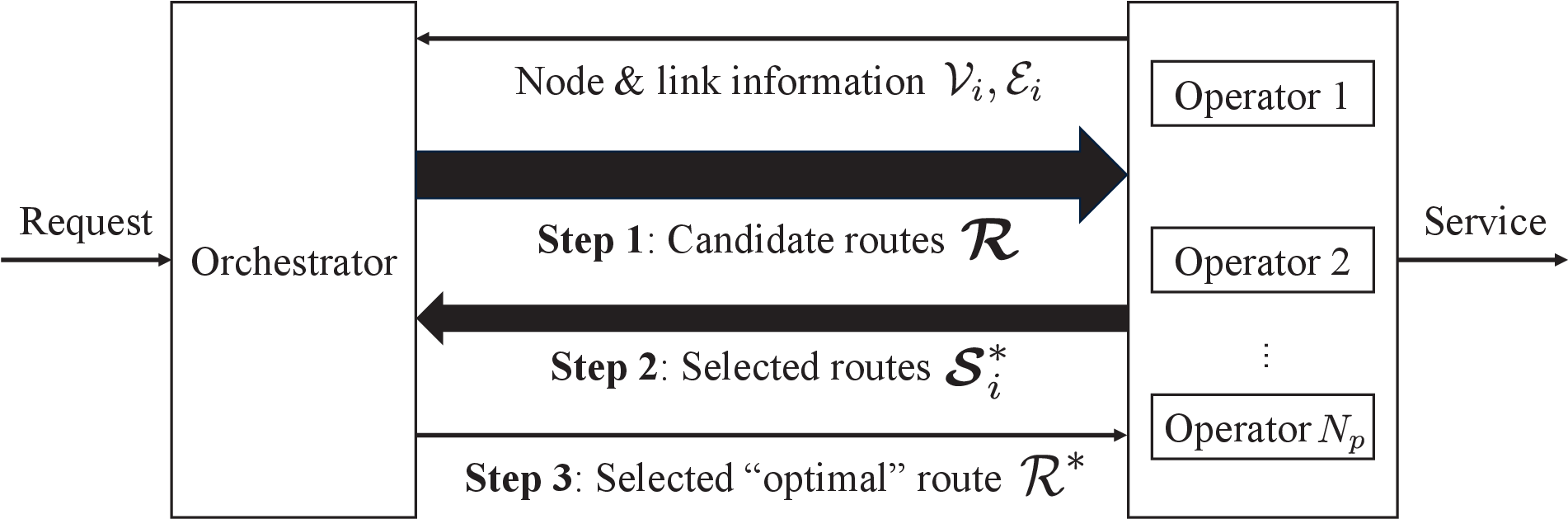}
            \label{fig:proposed_framework_a}
        }
    \end{minipage}
    \\ \smallskip
    \begin{minipage}[t]{0.98\hsize}
        \centering
        \subfloat[]{
            \includegraphics[width=0.55\columnwidth]{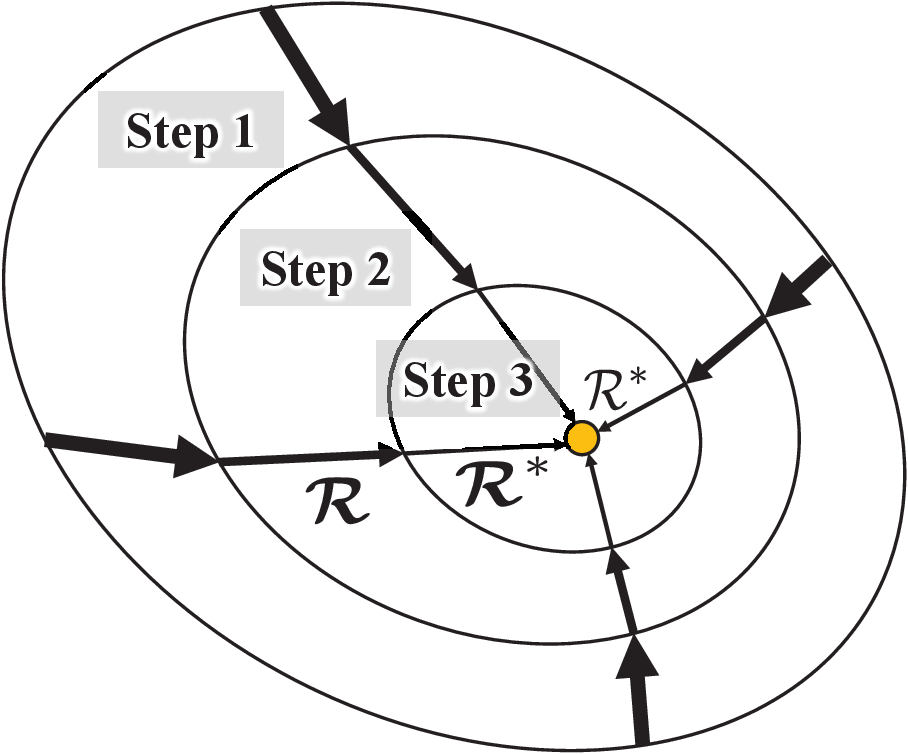}
            \label{fig:proposed_framework_b}
        }
    \end{minipage}
    \caption{The conceptual diagram of the proposed framework: (a) The information exchange between the orchestrator and the operators at each step. (b) The visualization of how the optimal route $\Rcal^{*}$ is gradually derived from the initial set of candidate routes $\bm{\Rcal}$ through the proposed three-step procedure. For simplicity, a set of route options is written by ellipses.}
    \label{fig:proposed_framework}
\end{figure}

To address this problem, we propose a weak-control-based NTN orchestration framework.
The original weak-control framework proposed in~\cite{Inoue_CSL19} incorporates human decision making into the feedback control loop, thereby granting humans freedom in determining the control input selection.
In this paper, we extend this concept to NTN orchestration by transferring part of this freedom to the operators, thereby enabling the consideration of their individual policies.
The key idea of the proposed framework is a three-step approach: (i) the orchestrator first narrows down the set of candidate routes, (ii) each operator then selects routes from this set according to its own policies, and (iii) the orchestrator performs the final optimization. 
In the first step, unlike conventional frameworks, the orchestrator does not solve an optimization problem but instead focuses on a constraint satisfaction problem. 
This ensures that the orchestrator only determines the feasible route set without enforcing a specific optimal route, which is left to the subsequent steps. 
Furthermore, the operators are not required to disclose their policies to the orchestrator, thereby preserving their privacy.

The conceptual diagram of the proposed framework is described in Fig.~\ref{fig:proposed_framework}.
Fig.~\ref{fig:proposed_framework_a} illustrates the information exchange between the orchestrator and the operators at each step.
First, the orchestrator receives the sets of nodes $\Vcal_{i}$ and links $\Ecal_{i}$ from each operator.
Based on this information, the output of Step~1 is the set of candidate routes generated by the orchestrator, the output of Step~2 is the set of routes selected by the operators according to their policies, and the output of Step~3 is the optimal route finally determined by the orchestrator.
Fig.~\ref{fig:proposed_framework_b} visualizes how the optimal route $\Rcal^{*}$ is gradually derived from the initial set of candidate routes $\bm{\Rcal}$ through the proposed three-step procedure.
In this figure, each ellipse represents a set of route options that can be selected either by the orchestrator or by the operators.

A list of notations and parameters defined in this paper is summarized in Table~\ref{table:notations_parameters}.

\begin{table}[tb]
\centering
\caption{List of notations and parameters in this paper.}
\footnotesize
\renewcommand{\arraystretch}{0.95}
    \begin{tabularx}{\columnwidth}{lX} \hline
        Notation & Description \\ \hline \hline
        \multicolumn{2}{@{}l}{[Basic Definitions]} \\
        $e,r$ & Link \\
        $\Rcal,\Pcal$ & Route, or portion of a route (i.e., set of links) \\
        $\bm{\Rcal},\bm{\Pcal}$ & Set of routes \\
        
        $N_{p},\Ncal_{p}$ & Number of operators and its index set \\

        $\Vcal_{i}$ & Set of NTN nodes owned by $i$-th operator \\   
        $\Ecal_{i}, \Ecal_{(i,j)}$ & Set of feasible links within $i$-th operator or between $i$-th and $j$-th operators \\            

        \hline
        \multicolumn{2}{@{}l}{[Policies, Cost Functions, and Constraints]} \\
        
        $\pi_{k}$ & Policy of orchestrator ($k=0$) or $i$-th operator ($k=i$) \\
        $J_{k}, \bar{J_{k}}$ & Cost function defined by $\pi_{k}$ and its upper bound \\
        $\Ccal_{k}$ & Set of constraints defined by $\pi_{k}$ \\
        
        \hline
        \multicolumn{2}{@{}l}{[Route Components]} \\
        $\bm{\Rcal}, \Rcal_{\ell}$ & Set of candidate routes obtained by orchestrator and $\ell$-th candidate route \\
        $N_{r},\Ncal_{r}$ & Number of candidate routes provided by orchestrator and its index set \\

        $\Rcal_{i,\ell}, r_{(i,j),\ell}$ & Set of links within $i$-th operator and link between $i$-th and $j$-th operators in $\ell$-th route \\

        $\Pcal_{i,\ell,k}, p_{(i,j),\ell,k}$ & Subset of links within $i$-th operator and link from $i$-th to $j$-th operator at $k$-th occurrence in $\ell$-th route \\

        $N_{i,\ell}$ & Number of occurrences of $i$-th operator in $\ell$-th route \\
        
        $\Qcal_{i,\ell,k}, \bar{\Qcal}_{i,\ell}$ & Subset of links in $\ell$-th route of $i$-th operator at $k$-th occurrence and its aggregated set ($\bar{\Qcal}_{i,\ell}=\{\Qcal_{i,\ell,k}\}_{k=1,2,\ldots,N_{i,\ell}}$) \\
        
        $\bm{\Scal}_{i}$ & Set of link subsets provided by orchestrator for $i$-th operator ($\bm{\Scal}_{i}=\{\bar{\Qcal}_{i,\ell}\}_{\ell\in\Ncal_{r}}$) \\

        $\bm{\Scal}_{i}^{*}, \Ncal_{i,r}^{*}$ & Set of selected routes of $i$-th operator ($\bm{\Scal}_{i}^{*}\subseteq\bm{\Scal}_{i}$) and its index set ($\Ncal_{i,r}^{*}\subseteq\Ncal_{r}$) \\
        
        $\bm{\Rcal}^{*}$ & Intersection of routes selected by all operators ($\bm{\Rcal}^{*}\subseteq\bm{\Rcal}$) \\
        $\Rcal^{*}$ & Optimal route determined by the proposed framework \\

        \hline
    \end{tabularx}
\label{table:notations_parameters}
\end{table}

\subsection{Procedure of Proposed Framework}
\label{section:proposed_orchestration_procedure}

In the following, we describe each step of the proposed framework in detail and define the corresponding problems to be solved.

\subsubsection{Step 1: Orchestrator Candidate Generation}

The goal of this step for the orchestrator is to identify a set of feasible candidate routes based on the feasible links $\Ecal_{i}$ and $\Ecal_{(i,j)}$, subject to the constraint $\Ccal_{0}$. 
This set is obtained by solving the following policy-constrained satisfaction problem.

\begin{problem}[Orchestrator's Constraint Satisfaction Problem]
    Given $\{\Vcal_{i}, \Ecal_{i}, \Ecal_{(i,j)}\}, i\in\Ncal_{p},j\neq i$, find $\bm{\Rcal}=\{\Rcal_{\ell}\}$ such that $\Rcal_{\ell}\in\Ccal_{0}$.
\label{problem:orchestrator_set_candidate_routes}
\end{problem}

We assume the orchestrator finds $N_{r}$ candidate routes, and a set of indices of candidate routes is defined by
\begin{align}
    \Ncal_{r} = \{1, 2, \ldots, N_{r}\}. \nonumber
\end{align}
Each route $\Rcal_{\ell}\in\bm{\Rcal}$ consists of links in each operator's network and links between the operators.
In this paper, we represent the relationship between the route and links in the following expression:
\begin{align}
    \Rcal_{\ell} = \bigcup_{i\in\Ncal_{p}} \Rcal_{i,\ell} \cup \bigcup_{i,j\in\Ncal_{p},i\neq j} r_{(i,j),\ell},
    \nonumber
\end{align}
where $\Rcal_{i,\ell}\subseteq\Ecal_{i}$ and $r_{(i,j),\ell}\in\Ecal_{(i,j)}$ represent a set of links related to the $\ell$-th candidate route in the $i$-th operator's network and a link across the $i$-th and $j$-th operators, respectively.
It is noted that:
\begin{itemize}
    \item Not all operators' networks are necessarily included in the candidate routes $\Rcal_{\ell}$, i.e., $\Rcal_{i,\ell}=\emptyset, r_{(i,j),\ell}=\emptyset, \exists i,j\in\Ncal_{p}, \exists\ell\in\Ncal_{r}$.
    \item Routes that pass through the same operator multiple times are also included.
    \item If there is only one node that passes through a certain operator, only inter-operator connections are defined for that operator, resulting in $\Rcal_{i,\ell}=\emptyset$.
\end{itemize}

\begin{figure}[tb]
	\centering
	\includegraphics[width=0.98\columnwidth]{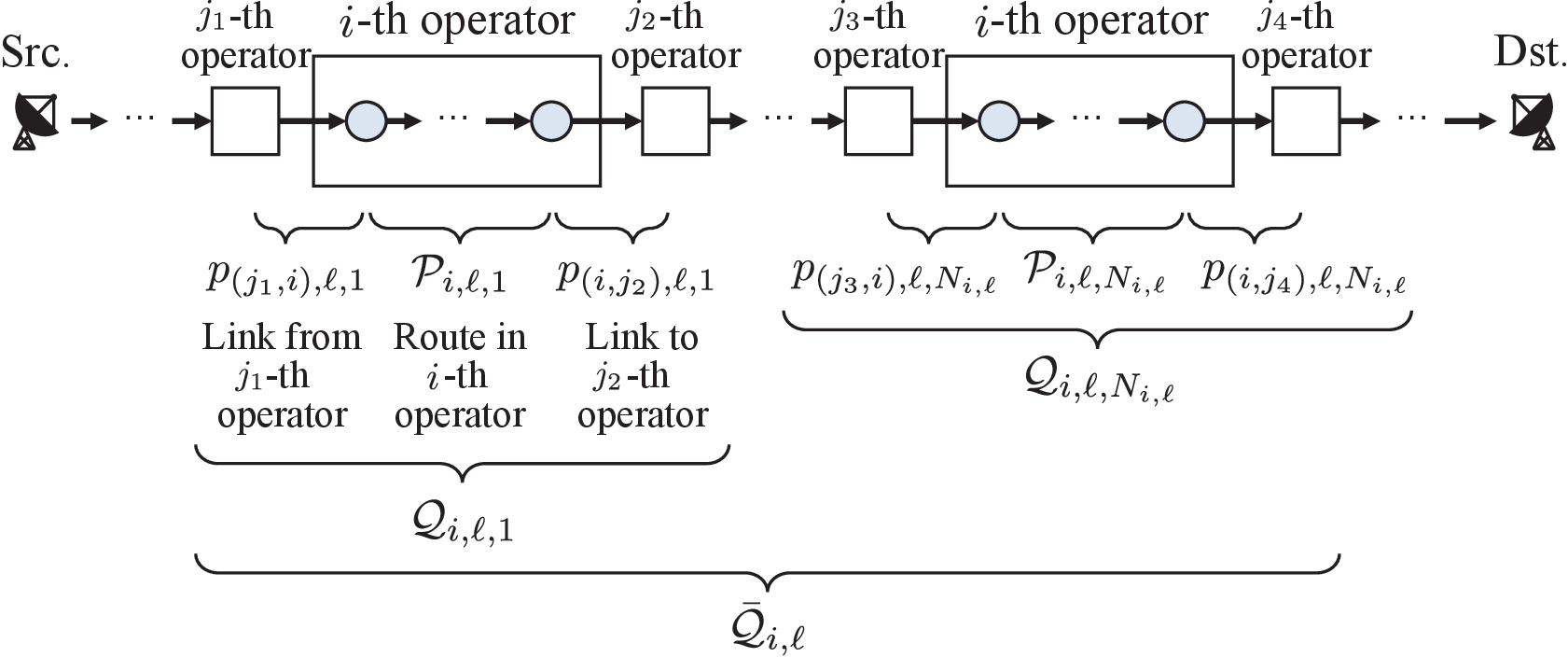}
	\caption{The $\ell$-th route candidate focusing on $i$-th operator. This figure shows an example of passing through the $i$-th operator twice.}
\label{fig:proposed_definition_route_set}
\end{figure}

We now define a set of feasible candidate routes to be notified to the $i$-th operator by the orchestrator, summarized in Fig.~\ref{fig:proposed_definition_route_set}.
The orchestrator divides the route candidates for each operator in the following form:
\begin{align}
    \Qcal_{i,\ell,k} = \{p_{(j_{1},i),\ell,k}, \Pcal_{i,\ell,k}, p_{(i,j_{2}),\ell,k}\},
    \label{eq:candidate_routes_for_operator_part}
\end{align}
where $p_{(j_{1},i),\ell,k}, \Pcal_{i,\ell,k},$ and $p_{(i,j_{2}),\ell,k}$ represent a link from $j_{1}$-th to $i$-th operator, a set of links within $i$-th operator, and a link from $i$-th to $j_{2}$-th operator, respectively.
Note that $j_{1}$-th and $j_{2}$-th can be the same.
A single route may pass through the same operator multiple times; therefore, we denote each occurrence of an operator along the route with an index $k$.
Here, we denote $N_{i,\ell}$ as the number of times the $\ell$-th route passes through the $i$-th operator's network.
For the $\ell$-th route, we define a set of $\Qcal_{i,\ell,k}$ in Eq.~(\ref{eq:candidate_routes_for_operator_part}) that are related to the $i$-th operator as
\begin{align}
    \bar{\Qcal}_{i,\ell} = \{\Qcal_{i,\ell,1}, \ldots, \Qcal_{i,\ell,N_{i,\ell}}\}.
    \label{eq:candidate_routes_for_operator}
\end{align}
As a result, the orchestrator provides the following set to each operator:
\begin{align}
    \bm{\Scal}_{i} = \{\bar{\Qcal}_{i,1}, \bar{\Qcal}_{i,2}, \ldots, \bar{\Qcal}_{i,N_{r}}\}.
    \label{eq:set_candidate_routes_for_operator}
\end{align}

\smallskip
\subsubsection{Step 2: Operator Policy Filtering}

Next, each operator tries to find a set of preferred candidate routes in its network according to their policies by solving the following constraint satisfaction problem:
\begin{problem}[Operator's Constraint Satisfaction Problem]
    Find $\bm{\Scal}_{i}^{*}=\{\bar{\Qcal}_{i,\ell}\} \subseteq \bm{\Scal}_{i}$ such that $\bar{\Qcal}_{i,\ell}\in\Ccal_{i}$.
\label{problem:operator_set_feasible_routes}
\end{problem}

The set $\bm{\Scal}_{i}^{*}$ contains the routes $\bar{\Qcal}_{i,\ell}$ that satisfy their conditions from among the candidates.
Here, a set containing the indices of the selected route is defined as $\Ncal_{i,r}^{*}\subseteq\Ncal_{r}$ for the $i$-th operator.
The operators can impose other constraints on this problem according to their other objectives if they have.
The resulting set is presented to the orchestrator.

\smallskip
\subsubsection{Step 3: Orchestrator Optimal Selection}

Finally, the orchestrator tries to calculate the intersection of the set of routes selected by each operator, represented by
\begin{align}
    \bm{\Rcal}^{*} = \left\{\Rcal_{\ell} \relmiddle| \ell \in \bigcap_{i\in\Ncal_{p}} \Ncal_{i,r}^{*} \right\} \subseteq \bm{\Rcal},
    \label{eq:set_orchestrator_feasible_routes}
\end{align}
where $\bm{\Rcal}^{*}$ is a subset of $\bm{\Rcal}$.
If $\bm{\Rcal}^{*}=\emptyset$, the orchestrator should manage to relax the constraint or negotiate with the operators to relax their constraints.
In that case, the orchestrator or operators relax the policies via the following updates: modifying the constraints as $\Ccal_{0} \leftarrow \Ccal'_{0}$ or $\Ccal_{i} \leftarrow \Ccal'_{i}$ and setting the evaluation function bounds to $\bar{J}_{0} + \epsilon_{0}$ or $\bar{J}_{i} + \epsilon_{i}$.
Here, $\Ccal'_{0}$ and $\Ccal'_{i}$ denote the relaxed constraints of the orchestrator and $i$-th operator, and $\epsilon$ denotes the relaxation margin.

The best route $\Rcal^{*}$ is obtained by solving the following optimization problem:
\begin{problem}[Route Optimization Problem]
    For a given $\bm{\Rcal}^{*}$ in Eq.~(\ref{eq:set_orchestrator_feasible_routes}), find $\Rcal^{*}$ that minimizes $J_{0}(\Rcal)$.
\label{problem:orchestator_best_route}
\end{problem}

\noindent
The obtained route $\Rcal^{*}$ is divided into the routes within each operator's network and communicated to the respective operators.

This method enables the design of a network that accommodates the policies of the orchestrator and each operator as much as possible.
In addition, compared to the case where the operators communicate directly with each other to determine the route, the orchestrator narrows down the candidates, which can also reduce each operator's operational cost.

We describe the algorithm of the proposed framework in Algorithm~\ref{algorithm:proposed_framework}.

\begin{figure}[tb]
    \begin{algorithm}[H]
        \caption{Proposed weak-control-based NTN orchestration}
        \label{algorithm:proposed_framework}
        \begin{algorithmic}[1]
        \REQUIRE $J_{0}$, $\bar{J}_{0}$, $\pi_{0}$, $J_{i}$, $\bar{J}_{i}$, $\pi_{i}$, $flag\_nonempty \leftarrow 0$ % Input
        \ENSURE Determined route $\Rcal^{*}$ % Output

        \FOR{Operator $i$ in $\Ncal_{p}$}
        \STATE Provide a set of nodes $\Vcal_{i}$ and the information of these nodes to the orchestrator
        \ENDFOR
        
        \WHILE{$flag\_nonempty = 0$}

        \STATE Calculate sets of links consisting of nodes of $i$-th operator $\Ecal_{i}$ and links consisting of nodes between operators $\Ecal_{(i,j)}$
        \STATE Solve \textbf{Problem~\ref{problem:orchestrator_set_candidate_routes}} to find a set of candidate routes $\bm{\Rcal}=\{\Rcal_{\ell}\}$.
        \STATE Calculate $\Qcal_{i,\ell,k}$ in Eq.~(\ref{eq:candidate_routes_for_operator_part}), $\bar{\Qcal}_{i,\ell}=\{\Qcal_{i,\ell,k}\}$ in Eq.~(\ref{eq:candidate_routes_for_operator}), and $\bm{\Scal}_{i}=\{\bar{\Qcal}_{i,\ell}\}$ in Eq.~(\ref{eq:set_candidate_routes_for_operator}) for each operator
        \STATE Provide $\bm{\Scal}_{i}$ to each operator
        
        \FOR{Operator $i$ in $\Ncal_{p}$}
        \STATE Solve \textbf{Problem~\ref{problem:operator_set_feasible_routes}} to find a set of preferable routes $\bm{\Scal}_{i}^{*}=\{\bar{\Qcal}_{i,\ell}\} \subseteq \bm{\Scal}_{i}$.
        \STATE Provide $\bm{\Scal}_{i}^{*}$ to the orchestrator
        \ENDFOR

        \STATE Calculate the intersection of the feasible routes $\bm{\Rcal}^{*}$ in Eq.(\ref{eq:set_orchestrator_feasible_routes})
        
        \IF{$\bm{\Rcal}^{*}\neq\emptyset$}
        \STATE Solve \textbf{Problem~\ref{problem:orchestator_best_route}} to find the best route $\Rcal^{*}$.
        \STATE $flag\_nonempty \leftarrow 1$
        \ELSE
        \STATE Relax the constraint of the orchestrator, $\bar{J}_{0} \leftarrow \bar{J}_{0} + \epsilon_{0}$ or each operator, $\bar{J}_{i} \leftarrow \bar{J}_{i} + \epsilon_{i}$
        \ENDIF
        
        \ENDWHILE
        \end{algorithmic}
    \end{algorithm}
\end{figure}

\begin{remark}
    The orchestrator puts different policies in the following two phases:
    \begin{itemize}
        \item Step 1; where the orchestrator calculates candidate route,
        \item Step 3; where the orchestrator receives the selected routes from each operator and calculates the final route.
    \end{itemize}
    In Step~1, policies $\pi_{\text{LH}}(n_{h}), \pi_{\text{LL}}(n_{l})$, and $\pi_{\text{LO}}(n_{o})$ can be applied.
    On the other hand, in Step~3, policies $\pi_{\text{MH}}, \pi_{\text{ML}}$, and $\pi_{\text{MO}}$ can be applied because the optimal route must be determined.
\end{remark}

\subsection{Illustrative Example of Proposed Orchestration: Two Operators Case}
\label{section:proposed_orchestration_example}

\begin{figure}[tb]
	\centering
	\includegraphics[width=0.97\columnwidth]{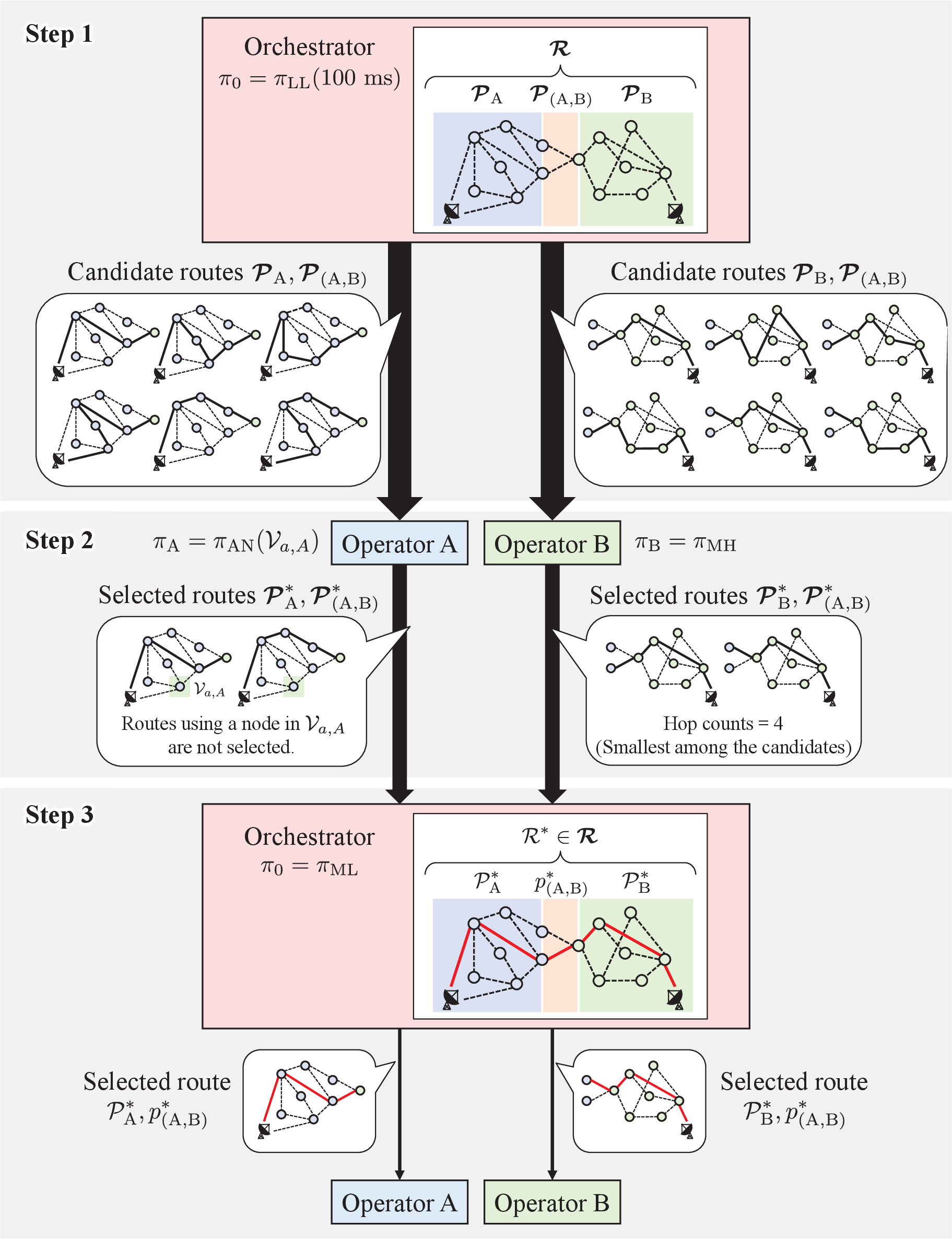}
	\caption{Conceptual diagram of the proposed orchestration framework in this example.}
\label{fig:proposed_framework_example}
\end{figure}

The conceptual diagram of the proposed orchestration framework in this example is shown in Fig.~\ref{fig:proposed_framework_example}.
In this example, the orchestrator's policy in Step~1 is defined as $\pi_{0}=\pi_{\text{LL}}(100~\text{ms})$, which ensures that the set of feasible routes excludes those exceeding 100~ms. 
The policies applied for the orchestrator in Step~3 and for Operators A and B are set as $\pi_{0}=\pi_{\text{ML}}$, $\pi_{\text{A}}=\pi_{\text{AN}}(\Vcal_{a,\text{A}})$, and $\pi_{\text{B}}=\pi_{\text{MH}}$, respectively, as in Section~\ref{section:centralized_orchestration_example}.

The procedure is described below and consists of three steps as shown in Section~\ref{section:proposed_orchestration_procedure}.
In Step~1, the orchestrator solves Problem~\ref{problem:orchestrator_set_candidate_routes} to identify sets of feasible routes $\bm{\Rcal}$ by using a constraint inequality $J_{0}\leq\text{100 ms}$.
The obtained set is divided into $\bm{\Pcal}_{\text{A}}$, $\bm{\Pcal}_{\text{B}}$, and $\bm{\Pcal}_{(\text{A},\text{B})}$, representing the sets of candidate routes for Operator~A, Operator~B, and the inter-operator connections, respectively.
Then, the orchestrator provides each set to Operators A and B.
In this example, the orchestrator gives six candidate routes to Operators A and B, respectively.

Then, each operator solves Problem~\ref{problem:operator_set_feasible_routes} and selects a feasible and preferable set of routes from the candidates: $\bm{\Pcal}_{\text{A}}^{*}\subseteq\bm{\Pcal}_{\text{A}}$ and $\bm{\Pcal}_{(\text{A},\text{B})}^{*}\subseteq\bm{\Pcal}_{(\text{A},\text{B})}$ for Operator~A and $\bm{\Pcal}_{\text{B}}^{*}\subseteq\bm{\Pcal}_{\text{B}}$ and $\bm{\Pcal}_{(\text{A},\text{B})}^{*}\subseteq\bm{\Pcal}_{(\text{A},\text{B})}$ for Operator~B according to their evaluation functions $J_{\text{A}}$ and $J_{\text{B}}$ based on their policies, respectively, and presents the set to the orchestrator in Step~2.

In this example, each operator returns two selected routes to the orchestrator. 
Finally, in Step~3, the orchestrator solves Problem~\ref{problem:orchestator_best_route} and determines a specific end-to-end route $\Rcal^{*}\in\bm{\Rcal}$. 
The obtained route is divided into $\Pcal^{*}_{\text{A}}\in\bm{\Pcal}_{\text{A}}^{*}$, $\Pcal^{*}_{\text{B}}\in\bm{\Pcal}_{\text{B}}^{*}$, and $p^{*}_{(\text{A},\text{B})}\in\bm{\Pcal}_{(\text{A},\text{B})}^{*}$. 
This solution is then communicated to the operators, who each establish their networks accordingly. 
This approach enables the construction of a network that aligns as closely as possible with the policies of both the orchestrator and the individual operators.

\subsection{Summary of Proposed NTN Orchestration}
\label{section:proposed_orchestration_summary}

\begin{figure}[tb]
	\centering
	\includegraphics[width=\columnwidth]{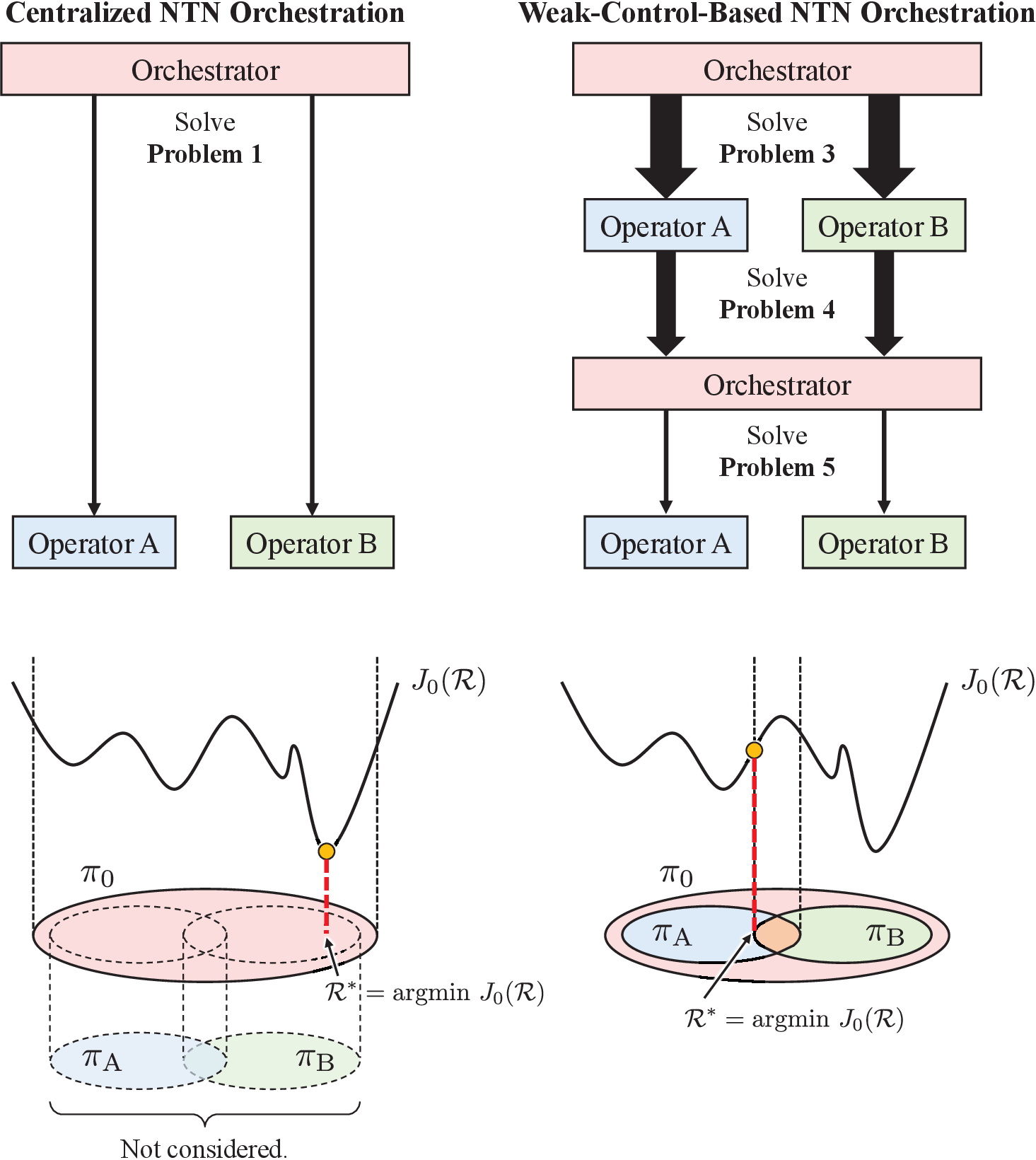}
	\caption{Comparison between the conventional centralized orchestration and the proposed weak-control-based framework using the two-operator example. For simplicity, the region of constraints is written by ellipses.}
\label{fig:proposed_framework_characterization}
\end{figure}

Fig.~\ref{fig:proposed_framework_characterization} compares the conventional centralized orchestration framework with the proposed weak-control-based framework, using the two-operator example introduced in Section~\ref{section:proposed_orchestration_example}. 
In the centralized framework, the orchestrator solves Problem~\ref{problem:conventional_orchestration} and provides the resulting route to the individual operators. 
This framework yields the solution $\Rcal^{*}$ that minimizes the orchestrator's cost function $J_{0}(\Rcal)$, without considering the policies of individual operators. 
By contrast, the proposed framework consists of three steps: the orchestrator solves Problem~\ref{problem:orchestrator_set_candidate_routes}, each operator solves Problem~\ref{problem:operator_set_feasible_routes}, and the orchestrator then solves Problem~\ref{problem:orchestator_best_route}. 
The resulting route does not necessarily minimize the orchestrator's cost function $J_{0}(\Rcal)$; however, it enables the determination of a suboptimal route that accounts for the operators' policies.

To clarify the theoretical foundation of our proposed method, we summarize the mapping between the original weak-control framework~\cite{Inoue_CSL19} and our proposed NTN orchestration context as follows:
\begin{itemize}
    \item ``Controller'' ($\mathcal{K}$): Represented by the orchestrator, which provides a set of feasible route options instead of a single optimal route.
    \item ``Set-valued Signal'' ($\mathcal{U}$): Represented by the set of candidate routes ($\mathcal{R}$) generated by the orchestrator in Step~1.
    \item  ``Decision Maker'' ($\mathcal{H}$): Represented by the collective of NTN operators and the orchestrator. Although the original framework assumes a single decision maker, our context interprets the combined selection process in Steps~2 and 3, based on policies $\pi_{i}$ and $\pi_{0}$, as the decision-making behavior.
    \item ``Plant'' ($P$): Represented by the actual NTN infrastructure established and managed by the operators.
\end{itemize}
By adopting this theoretical perspective, our framework effectively balances global network performance by the orchestrator with the local autonomy of individual NTN operators.

Reflecting the scale and dynamic nature of NTN, the computational feasibility of the proposed framework is a key consideration.
In a standard shortest path problem, algorithms such as Dijkstra's can solve it in $O(|\Vcal|^{2})$, and this can be improved to $O(|\Ecal|\log|\Vcal|)$ or $O(|\Ecal|+|\Vcal|\log|V|)$ under appropriate conditions~\cite[Chapter~4]{Ahuja_93}.
In contrast, in the problem considered in this paper, Problem~1 becomes a mixed-integer programming problem and is therefore NP-hard.
In particular, Problem~3, which is formulated as a constraint satisfaction problem, can grow exponentially depending on the network characteristics and policies; for example, in a graph with average degree $\mu$, the number of paths with at most $n_{h}$ hops (as assumed under the hop-limit policy $\pi_{\text{LH}}(n_{h})$) can be $O(\mu^{n_{h}})$, and in a dense graph it can be $O(|\Vcal|^{n_{h}})$~\cite[Chapter~5]{Russell_20}.
This indicates that it is impractical for the orchestrator to calculate \textit{all} possible routes, and from an implementation perspective, it is reasonable to impose an upper limit on the number of candidate routes $N_{r}$ calculated by the orchestrator.

To manage this computational complexity, the orchestrator can employ efficient algorithms, such as k-shortest path (e.g., Yen's algorithm~\cite{Hershberger_TALG07}), Mixed-Integer Programming (MIP)-based solution pooling (e.g., Branch-and-Cut~\cite{Stubbs_MP99}), or heuristic searches (e.g., Beam Search~\cite[Chapter~3]{Russell_20}), to identify only the top-$N_r$ candidate routes optimized for the objective (e.g., latency and hop count).
We note that this limit $N_{r}$ is an implementation parameter for scalability, not a restrictive policy.

Furthermore, to account for computation time, our framework can adopt a predictive strategy that leverages the deterministic nature of satellite orbits.
Instead of using the satellite network topology snapshot at the current time $t_{\text{now}}$, the orchestrator predicts the topology at the future route establishment time $t_{\text{est}} = t_{\text{now}} + T_{\text{pred}}$, where $T_{\text{pred}}$ represents the prediction time horizon accounting for computation and signaling delays.
However, inherent errors in orbital prediction may cause differences between the actual physical state and the logical topology computed by the orchestrator, potentially preventing candidate routes from being physically established.
To address this, our proposed framework employs a multi-layered robustness strategy.
Minor prediction errors are compensated for by \textit{physical control mechanisms} (e.g., satellite attitude control and fine pointing), while \textit{route diversity} allows operators to select alternative routes from the presented route set if a specific link becomes unavailable.
Finally, for significant disruptions that make most of the candidate set infeasible, the system triggers \textit{re-orchestration} to dynamically find new feasible routes.
This re-orchestration mechanism is also effective for unexpected link failures due to stochastic factors (e.g., weather conditions) or sudden traffic surges.

From the perspective of control signaling, it is worth noting that the impact of the overhead in the proposed framework is relatively small.
Signaling between the orchestrator and operators is conducted over terrestrial networks, thereby avoiding the consumption of limited satellite communication resources.
Furthermore, the traffic load is minimized by exchanging only abstract information (e.g., sets of candidate routes and selected route indices) at a low frequency (e.g., every few minutes or longer), unlike the millisecond- or second-level signaling required for per-packet operations.

\section{Numerical Simulation}
\label{section:simulation}

In this section, we conduct five types of numerical simulations to evaluate the effectiveness of our proposed orchestration framework.
In order to show the characteristics of operator coordination, routes that can be transmitted only by a single operator shall be excluded from the candidate routes presented by the orchestrator.

\subsection{Effects of Proposed Framework in Two Operators Case}
\label{section:simulation1}

First, we verify the effectiveness of our proposed orchestration framework.
In this simulation, as shown in Table~\ref{table:simulation_operators_two_operators}, we assume the presence of the orchestrator and two NTN operators, i.e., $N_{p}=2$.
A total of ten LEO orbital planes, each containing ten satellites (100 satellites in total), are configured.
The orbital parameters of these LEO satellites are listed in Table~\ref{table:simulation_parameters}.
These orbital planes are alternately owned by the two operators, with each operator possessing every other orbital plane (five orbital planes and 50 LEO satellites each).
The ``User'', located in New York, U.S. (coordinates: 40.68939$^{\circ}$N, 74.04453$^{\circ}$W), requests a communication link to a physically located ``DN'' (Data Network) in Tokyo, Japan (coordinates: 35.710076$^{\circ}$N, 139.489154$^{\circ}$E).
An optical ground station (OGS) that communicates with satellites is placed at the same location as the DN and is represented as ``OGS'' (coordinates: 35.710076$^{\circ}$N, 139.489154$^{\circ}$E).
We assume that the User, OGS, and DN do not belong to any operator.
This means that any operator's satellite can establish a connection with the OGS as long as the link conditions are satisfied.
The simulation in this subsection uses the link parameters of LEO satellites and OGS listed in Table~\ref{table:simulation_link_parameters} in Appendix~\ref{section:link_parameters}.
We first evaluate the results of a single time step at 2024/12/15 00:00 UTC, followed by an analysis of the dynamic and time-varying performance over a 60-minute duration.

The policies of the orchestrator and each NTN operator are as follows:
\begin{itemize}
  \item Orchestrator: 
    \begin{itemize}
        \item Step~1: $\pi_{\text{LH}}(10)$, to prefer routes with the hop count less than or equal to ten;
        \item Step~3: $\pi_{\text{ML}}$, to minimize traffic latency;
    \end{itemize}
  \item Operator A: $\pi_{\text{AN}}(\text{LEO-A-34}, \text{LEO-A-43})$, to avoid using specific satellites, LEO-A-34 and LEO-A-43, which are located around the source node in its network at the targeted time step;
  \item Operator B: $\pi_{\text{MH}}$, to minimize the hop count of satellites routed in its network.
\end{itemize}

\begin{table}[tb]
    \centering
    \caption{Simulation settings of the orchestrator and operators, their nodes, and policies in this scenario.}
    \scalebox{1}{
    \begin{tabular}{llcl} \hline 
        Parameter & Value \\ \hline \hline
        \textbf{Orchestrator} \\
        \hline
        \multirow{2}{*}{Policy} & $\pi_{0}=\pi_{\text{LH}}(10)$ in Step~1 \\
        & $\pi_{0}=\pi_{\text{ML}}$ in Step~3 \\
        \hline
        \textbf{Operator A} \\
        \hline
        NTN node & LEO-A-1 - LEO-A-50 \\
        \# of NTN nodes & 50 \\
        Policy & $\pi_{\text{A}}=\pi_{\text{AN}}(\text{LEO-A-34}, \text{LEO-A-43})$ \\
        \hline
        \textbf{Operator B} \\
        \hline
        NTN node & LEO-B-1 - LEO-B-50 \\
        \# of NTN nodes & 50 \\
        Policy & $\pi_{\text{B}}=\pi_{\text{MH}}$ \\
    \hline
    \end{tabular}
    }
\label{table:simulation_operators_two_operators}
\end{table}

\begin{table}[tb]
    \centering
    \caption{Satellite orbit parameters.}
    \scalebox{1}{
        \begin{tabular}{ll} \hline
            Parameter & Value \\ \hline \hline
            \# of satellites in total & 100 \\
            ~~\# of planes & 10 \\
            ~~\# of satellites in each plane & 10 \\
            Altitude & 1,000~km \\
            Inclination & 55~deg \\
            Eccentricity & 0~deg \\ 
            Difference of RAAN$^{*}$ with adjacent orbit & 36~deg \\ 
            \hline
            $^{*}$RAAN: Right Ascension of Ascending Node
        \end{tabular}
    }
\label{table:simulation_parameters}
\end{table}

\begin{table*}[tb]
  \centering
  \caption{The relationship between the number of candidate routes presented by the orchestrator, the number of routes selected by the operators, and the result of the obtained route in the scenario in Section~\ref{section:simulation1}. In this table, ``\text{LEO-A-*}'' is denoted by ``\text{A-*}'' for simplicity.}
  \scalebox{0.89}{
    \begin{tabular}{cccccccclcc} \hline 
      \makecell{ID} & 
      \makecell{$\pi_{0}$ \\ (Step~1)} & 
      \makecell{$\pi_{\text{A}}$} & 
      \makecell{$\pi_{\text{B}}$} & 
      \makecell{$N_{r}$} & 
      \makecell{$|\bm{\Scal}^{*}_{\text{A}}|$} & 
      \makecell{$|\bm{\Scal}^{*}_{\text{B}}|$} & 
      \makecell{$|\bm{\Rcal}^{*}|$} & 
      \makecell[l]{Resulting route} & 
      \makecell{$n^{*}_{h}$} & 
      \makecell{$\ell^{*}$} \\ \hline \hline

    \hline
    \multicolumn{11}{l}{\textbf{Centralized Framework}} \\
    \hline
      
      C-1 & - & - & - & - & - & - & - & 
      $\text{User} \rightarrow \underbrace{\red{\text{A-43}}}_{\text{Operator~A}} \rightarrow \underbrace{\text{B-24} \rightarrow \text{B-23}}_{\text{Operator~B}} \rightarrow \text{OGS} \rightarrow \text{DN}$ & 
      5 & 45.54~ms \\

    \hline
    \multicolumn{11}{l}{\textbf{Proposed Framework}} \\
    \hline
      
      P-1 & $\pi_{\text{LH}}(10)$ & $\pi_{\text{AN}}(\text{A-34}, \text{A-43})$ & $\pi_{\text{MH}}$ &
      1325 & 829 & 30 & 11 & 
      $\text{User} \rightarrow \underbrace{\text{B-25}}_{\text{Operator~B}} \rightarrow \underbrace{\text{A-25} \rightarrow \text{A-24} \rightarrow \text{A-23}}_{\text{Operator~A}} \rightarrow \text{OGS} \rightarrow \text{DN}$ & 
      6 & 65.49~ms \\
      
      P-2 & $\pi_{\text{LH}}(10)$ & $\pi_{\text{AN}}(\text{A-34}, \text{A-43}) \land \pi_{\text{MH}}$ &
      $\pi_{\phi}$ & 1325 & 27 & 1325 & 27 & 
      $\text{User} \rightarrow \underbrace{\text{B-34} \rightarrow \text{B-33} \rightarrow \text{B-32}}_{\text{Operator~B}} \rightarrow \underbrace{\text{A-32}}_{\text{Operator~A}} \rightarrow \text{OGS} \rightarrow \text{DN}$ & 
      6 & 56.95~ms \\
      
      \hline
    \end{tabular}
  }
  \label{table:simulation_results_two_operators}
\end{table*}

\begin{figure}[tb]
    \centering
    \begin{minipage}[t]{0.32\hsize}
        \centering
        \subfloat[]{
            \includegraphics[width=0.93\columnwidth]{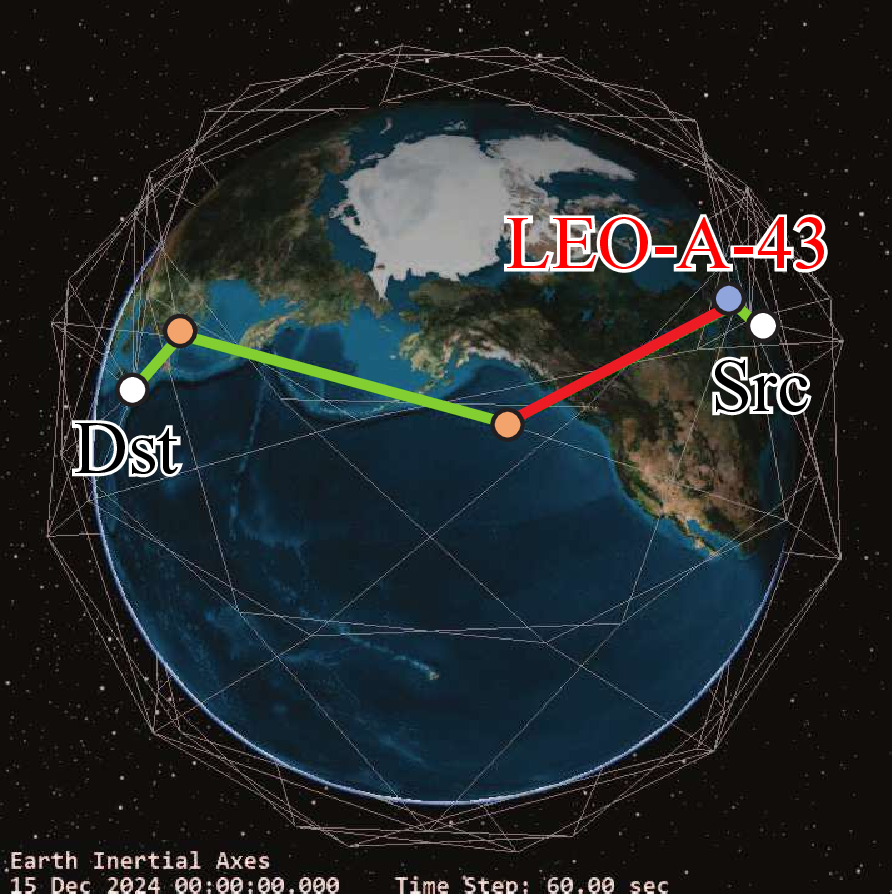}
            \label{fig:result_route_centralized}
        }
    \end{minipage}
    \hfill
    \begin{minipage}[t]{0.32\hsize}
        \centering
        \subfloat[]{
            \includegraphics[width=0.93\columnwidth]{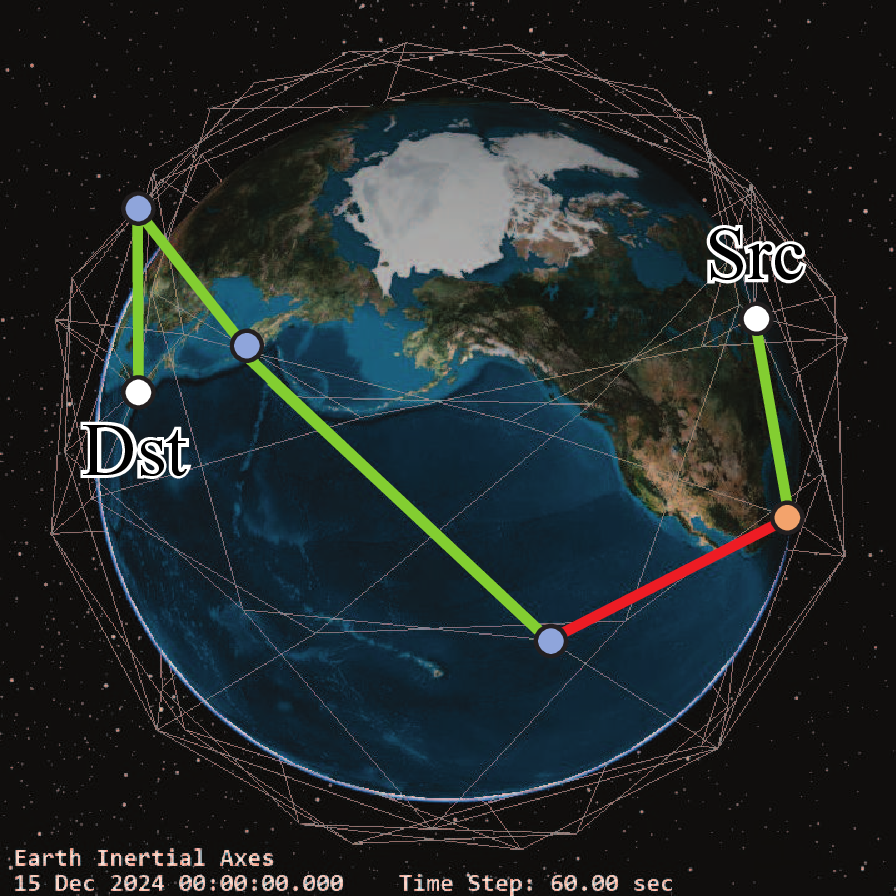}
            \label{fig:result_route_proposed_original}
        }
    \end{minipage}
    \hfill
    \begin{minipage}[t]{0.32\hsize}
        \centering
        \subfloat[]{
            \includegraphics[width=0.93\columnwidth]{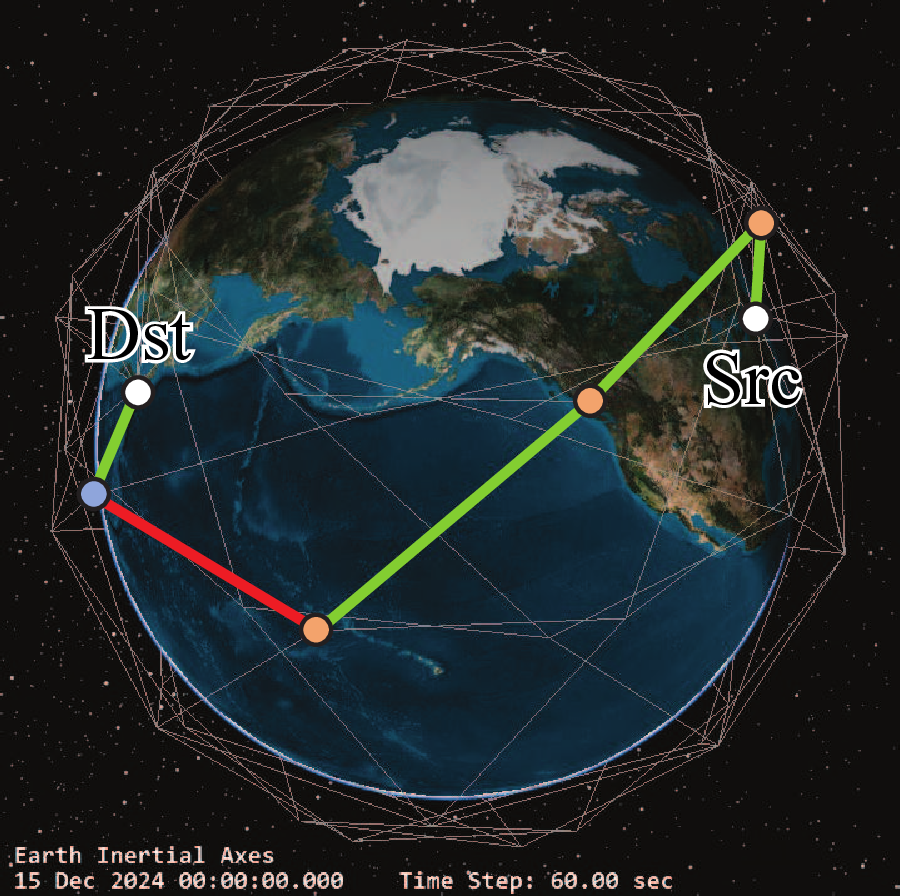}
            \label{fig:result_route_proposed_relaxed}
        }
    \end{minipage}
    \caption{The graphical results of the selected route by each framework: (a) Centralized framework, (b) Proposed framework, and (c) Proposed framework without the policy of Operator~B. Blue and red nodes belong to Operators~A and B, respectively. Green and red lines represent the route, and red lines represent the inter-operator connections especially. These figures were created using the Systems Tool Kit (STK).}
    \label{fig:results_route_stk}
\end{figure}

We compare the route obtained by the centralized orchestration framework with that obtained by the proposed framework.
In the following, $n^{*}_{h}$ and $\ell^{*}$ denote the hop count and latency of the obtained route.
First, we discuss the case of the centralized orchestration framework, and the route obtained by the framework is shown in C-1 of Table~\ref{table:simulation_results_two_operators}, where the latency was 45.54~ms.
The resulting route from the source to the destination is visualized in Fig.~\ref {fig:result_route_centralized}.
However, this route does not take into account Operator~A's policy of ``Avoid using specific satellites in its network'' and Operator~B's policy of ``Minimize the hop count of satellites routed in its network.''.
In fact, this route passes through LEO-A-43 in Operator~A's network.

In the proposed framework, the orchestrator presents candidate routes for each operator, and the operator selects preferable routes from among them.
First, the orchestrator calculated all routes based on the policy $\pi_{0}=\pi_{\text{LH}}(10)$, and the number of candidate routes was $N_{r}=1325$.
The orchestrator calculates a set of candidate routes $\bm{\Scal}_{i}$ in Eq.~(\ref{eq:set_candidate_routes_for_operator}) for each operator and presents them to the operators, and each operator selects a set of routes that satisfy its policy.
In this scenario, the number of routes satisfying the requirements of Operators~A and B were $|\bm{\Scal}^{*}_{\text{A}}|=829$ and $|\bm{\Scal}^{*}_{\text{B}}|=30$, respectively.
Finally, each operator sends the satisfied routes to the orchestrator, and the orchestrator selects one specific route that complies with its objective.
Note that a total of $|\bm{\Rcal}^{*}| = 11$ routes satisfied the requirements of both Operators~A and B, as determined by the intersection of their selected route sets.
We show the result in P-1 of Table~\ref{table:simulation_results_two_operators}, and the resulting route is visualized in Fig.~\ref {fig:result_route_proposed_original}.
The hop count $n^{*}_{h}$ was six and the latency $\ell^{*}$ was 65.49~ms, larger and longer than the previous result, respectively
However, regarding the operator's policy, this route does not pass through LEO-A-43 in Operator~A's network, and the hop count for Operator~B is minimized by passing through only LEO-B-25; therefore, each operator's policies are considered in the proposed framework.

The policies set by each operator strongly influence the obtained results.
In fact, if Operator~A also adopts the policy $\pi_{\text{MH}}$, as Operator~B does, there is no common set of routes chosen by Operators~A and B, preventing the orchestrator from determining a single route.
In such cases, the orchestrator or operators must relax their policies to obtain a feasible route, as described in Section~\ref{section:proposed_framework} and Algorithm~\ref{algorithm:proposed_framework}.
A detailed evaluation of this iterative negotiation and relaxation process is presented in Section~\ref{section:simulation1_negotiation}.
For example, we assume that Operator~B relaxes its policy from $\pi_{\text{MH}}$ to $\pi_{\phi}$, which means no policy.
In this case, the number of routes satisfying the policy has decreased from 829 to 27 for Operator~A with the added constraint and increased from 30 to 1325 for Operator~B with the relaxed constraint, as shown in P-2 of Table~\ref{table:simulation_results_two_operators}.
The resulting route was different from the previous condition visualized in Fig.~\ref {fig:result_route_proposed_relaxed}, and the total latency was 56.95~ms.

\begin{figure}[tb]
    \centering
    \begin{minipage}[t]{0.98\hsize}
        \centering
        \subfloat[]{
            \includegraphics[width=0.9\columnwidth]{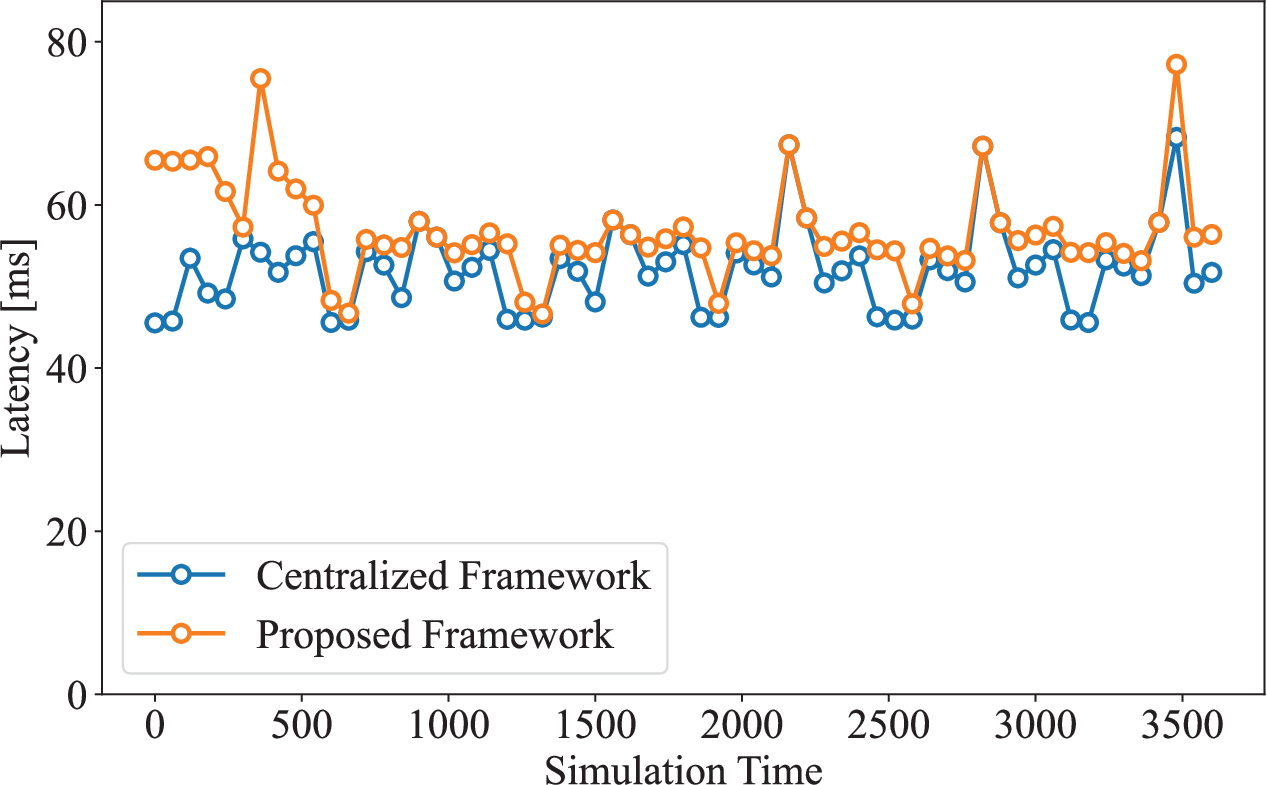}
            \label{fig:result_latency_time_series}
        }
    \end{minipage}
    \\ \smallskip
    \begin{minipage}[t]{0.98\hsize}
        \centering
        \subfloat[]{
            \includegraphics[width=0.9\columnwidth]{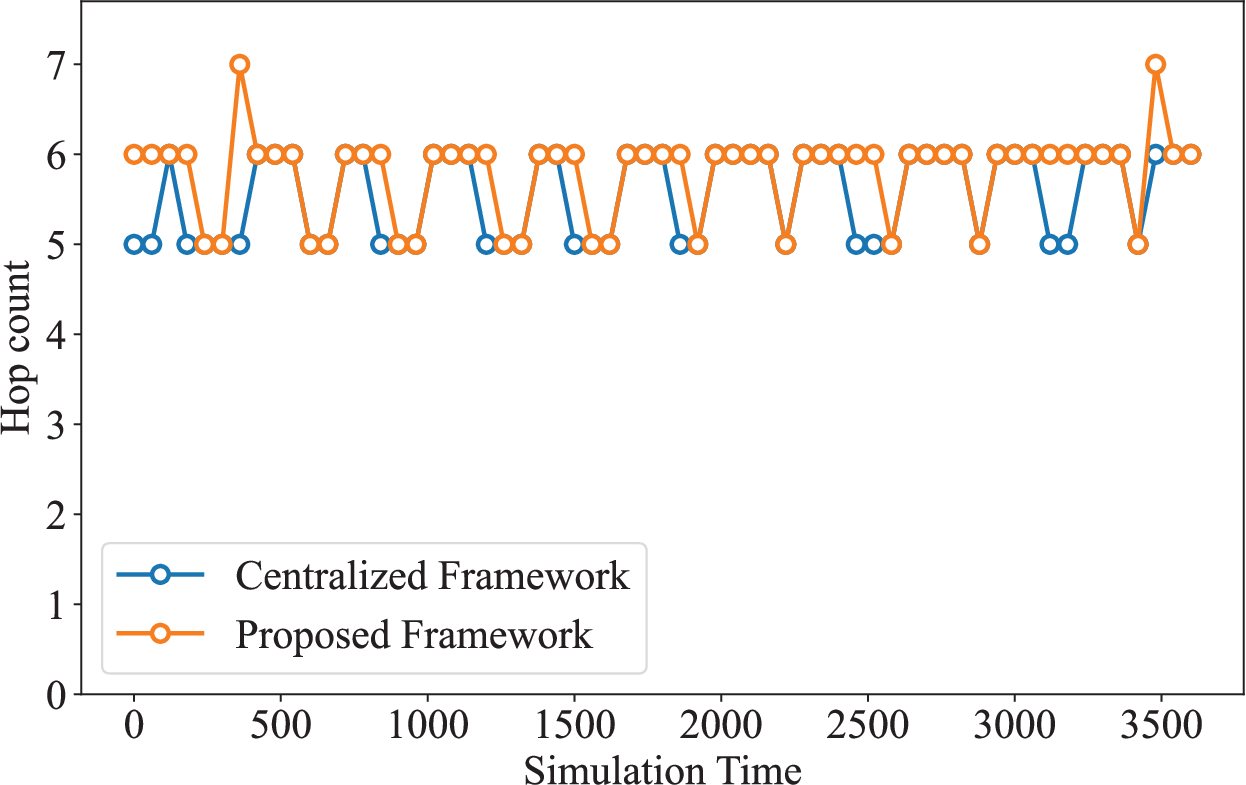}
            \label{fig:result_numhops_time_series}
        }
    \end{minipage}
    \caption{Time-varying results of (a) end-to-end latency and (b) hop count over 60 minutes with a 60-second time step. The blue and orange lines represent the results of the centralized framework (optimal route without policy constraints) and the proposed framework (with operator policies), respectively.}
    \label{fig:result_latency_numhop}
\end{figure}

Finally, we investigate the dynamic and time-varying performance of the proposed method.
We set a continuous 60-minute duration (from 2024/12/15 00:00 to 01:00 UTC) with a 60-second time step, resulting in a total of 61 time steps.
The orchestrator and operator policies are configured as shown in Table~\ref{table:simulation_operators_two_operators}.
Fig.~\ref{fig:result_latency_numhop} illustrates the time-varying latency and hop count results, comparing the centralized (blue line) and proposed frameworks (orange line).
Throughout this simulation period, the proposed framework successfully established feasible routes at all time steps, demonstrating its ability to adapt to dynamic topology changes in LEO networks.
As shown in Fig.~\ref{fig:result_latency_time_series}, the latency of the proposed framework is consistently higher than or equal to that of the centralized approach.
Similarly, in Fig.~\ref{fig:result_numhops_time_series}, the proposed framework typically requires the same number of hops, or occasionally one or two additional hops, compared to the centralized case to bypass restricted nodes.
This performance gap represents the dynamic ``price of autonomy'' caused by respecting operator policies, such as node avoidance (which will be discussed in detail in Section~\ref{section:simulation1_optimality_gap}).
These results confirm that the proposed framework ensures continuous service availability and stability even in a dynamic environment, effectively balancing the orchestrator's global performance with operator autonomy.

\subsection{Evaluation of Negotiation Mechanism}
\label{section:simulation1_negotiation}

In this subsection, we evaluate the effectiveness of the iterative negotiation mechanism introduced in Section~\ref{section:proposed_orchestration_procedure} and Algorithm~\ref{algorithm:proposed_framework}.
To simulate a policy-conflict scenario, we initially configured the operators with strict policies: Operator~A as $\pi_{\text{AN}}(\text{LEO-A-34}, \text{LEO-A-43}) \land \pi_{\text{MH}}$ and Operator~B as $\pi_{\text{MH}}$.
The orchestrator initially set the hop count limit $\pi_{\text{LH}}(8)$ and latency limit $\pi_{\text{LL}}(50~\text{ms})$.
Starting from this conflict state, we simulated the negotiation process where the orchestrator and operators incrementally relax their constraints to reach a feasible solution, and analyzed multiple relaxation patterns.

It is important to note that the relaxation strategy differs for each role, with specific trade-offs.
The orchestrator relaxes global constraints (e.g., latency) to expand the search space when candidates are insufficient, whereas operators relax local preferences (e.g., node avoidance) to accept a compromise among the presented candidates.
These parameters must be carefully selected to avoid excessive relaxation, which could violate the minimum operational requirements set by the orchestrator or significantly deviate from the original operator policies.
Therefore, in this simulation, we used fixed step sizes for the orchestrator ($\epsilon_{0}=1~\text{hop}$ or $10~\text{ms}$) and discrete relaxation steps for operators (e.g., removing node avoidance constraints or specific policies) to ensure efficient convergence while maintaining solution quality.

\begin{figure}[tb]
	\centering
	\includegraphics[width=0.95\columnwidth]{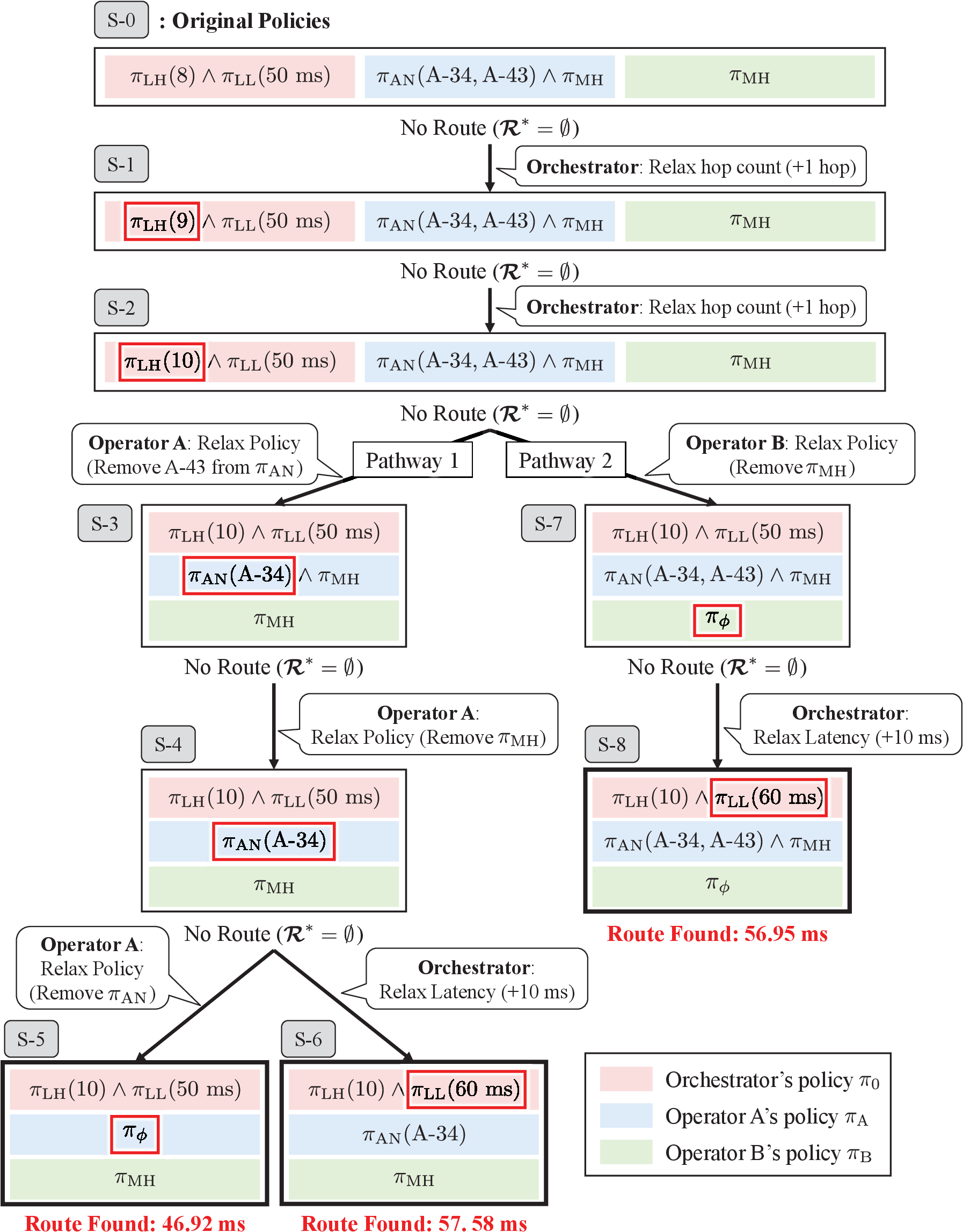}
	\caption{Flowchart of the iterative negotiation process starting from the conflict scenario, illustrating the incremental relaxation pathways (S-0 to S-8) taken by the orchestrator and operators to resolve the conflict. In this figure, ``\text{LEO-A-*}'' is denoted by ``\text{A-*}'' for simplicity.}
\label{fig:result_negotiation_relaxation_flow}
\end{figure}

Fig.~\ref{fig:result_negotiation_relaxation_flow} illustrates the flowchart of the conflict resolution process via negotiation.
Table~\ref{table:result_negotiation_relaxation} presents the number of candidate, selected, and common feasible routes ($N_{r}, |\bm{\Scal}^{*}_{\text{A}}|, |\bm{\Scal}^{*}_{\text{B}}|, |\bm{\Rcal}^{*}|$) and the network performance ($n^{*}_{h}, \ell^{*}$) obtained at each negotiation step, corresponding to the setting IDs shown in Fig.~\ref{fig:result_negotiation_relaxation_flow}.
As illustrated in the flowchart, the process begins with a common relaxation phase, where the orchestrator incrementally relaxes the hop count constraint from $\pi_{\text{LH}}(8)$ to $\pi_{\text{LH}}(10)$ (S-0 to S-2, $\epsilon_{0}=1$).
However, this alone does not resolve the conflict.
Subsequently, the negotiation process bifurcates into two main branches based on which operator relaxes their policy first:
\begin{itemize}
    \item \textbf{Pathway 1:} Relaxation Initiated by Operator~A (Left Branch).
    Operator~A progressively relaxes its constraints: first removing the avoidance of node LEO-A-43 (S-3), and then removing the hop-minimization policy $\pi_{\text{MH}}$ (S-4).
    From this state, two directions are observed:
    \begin{itemize}
        \item Operator~A Relaxation (S-5):
        If Operator~A further removes the remaining node-avoidance policy $\pi_{\text{AN}}$ (fully relaxing to $\pi_{\phi}$), a route is established with 46.92~ms latency.
        \item Orchestrator Relaxation (S-6):
        If Operator~A maintains the constraint $\pi_{\text{AN}}$, the orchestrator resolves the remaining conflict by relaxing the latency requirement from $\pi_{\text{LL}}(50~\text{ms})$ to $\pi_{\text{LL}}(60~\text{ms})$ ($\epsilon_{0}=10~\text{ms}$).
        This results in a feasible route with 57.58~ms latency.
    \end{itemize}
    \item \textbf{Pathway 2:} Relaxation Initiated by Operator~B (Right Branch).
    Operator~B relaxes its policy by removing $\pi_{\text{MH}}$ (S-7).
    Since a feasible route is still unavailable under the latency constraint, the orchestrator contributes to the resolution by relaxing the latency requirement from $\pi_{\text{LL}}(50~\text{ms})$ to $\pi_{\text{LL}}(60~\text{ms})$ ($\epsilon_{0}=10~\text{ms}$). 
    This collaborative relaxation establishes a route with 56.95~ms latency (S-8).
\end{itemize}
These results demonstrate that the proposed framework can dynamically find a ``compromise'' solution aligned with the specific policy priorities of the orchestrator and the operators.
From a practical perspective, while this mechanism ensures technical feasibility, incentive or compensation schemes would be necessary to motivate operators to accept such policy relaxations.

\begin{table}[tb]
  \centering
  \caption{Number of candidate, selected, and common feasible routes and network performance obtained at each negotiation step corresponding to the setting IDs shown in Fig.~\ref{fig:result_negotiation_relaxation_flow}. The notation ``-'' is used for $n^{*}_{h}$ and $\ell^{*}$ when no feasible route is found.}
  \scalebox{1}{
    \begin{tabular}{ccccccccc} \hline
      \makecell{Setting ID \\ (Fig.~\ref{fig:result_negotiation_relaxation_flow})} & 
      \makecell{$N_{r}$} & 
      \makecell{$|\bm{\Scal}^{*}_{\text{A}}|$} & 
      \makecell{$|\bm{\Scal}^{*}_{\text{B}}|$} & 
      \makecell{$|\bm{\Rcal}^{*}|$} & 
      \makecell{$n^{*}_{h}$} & 
      \makecell{$\ell^{*}$} \\ \hline \hline
      S-0 & 2 & 0 & 1 & 0 & - & - \\
      S-1 & 2 & 0 & 1 & 0 & - & - \\
      S-2 & 2 & 0 & 1 & 0 & - & - \\
      S-3 & 2 & 1 & 1 & 0 & - & - \\
      S-4 & 2 & 1 & 1 & 0 & - & - \\
      S-5 & 2 & 2 & 1 & 1 & 5 & 46.92~ms \\
      S-6 & 11 & 3 & 11 & 3 & 6 & 57.58~ms \\
      S-7 & 2 & 0 & 2 & 0 & - & - \\
      S-8 & 11 & 2 & 11 & 2 & 6 & 56.95~ms \\
      \hline
    \end{tabular}
  }
  \label{table:result_negotiation_relaxation}
\end{table}

\subsection{Analysis of Operator Autonomy}
\label{section:simulation1_optimality_gap}

As we have already investigated in the previous simulations, strict policies ensure operator autonomy but can lead to infeasible routing scenarios or performance degradation.
To quantify the trade-off between operator autonomy and global performance, we analyze two metrics:
\begin{itemize}
    \item \textit{Feasibility Rate:}
    The probability that a feasible route exists under the policy constraints in the proposed framework.
    \item \textit{Conditional Optimality Gap:}
    The average latency degradation compared to the centralized optimal route, calculated only for the trials where a feasible route was found.
    It is given by
    \begin{align}
        \dfrac{1}{N_{s}} \sum_{i=1}^{N_{s}}\dfrac{\ell^{*}_{\text{p},i} - \ell^{*}_{\text{c},i}}{\ell^{*}_{\text{c},i}}, \nonumber
    \end{align}
    where $N_{s}$ denotes the number of trials where a feasible route was found, and $\ell^{*}_{\text{p},i}$ and $\ell^{*}_{\text{c},i}$ denote the latency obtained via the proposed and centralized frameworks in the $i$-th trial, respectively.
    Note that when the feasibility rate is low, this metric is derived from a small number of samples.
\end{itemize}

\begin{figure}[tb]
	\centering
	\includegraphics[width=0.95\columnwidth]{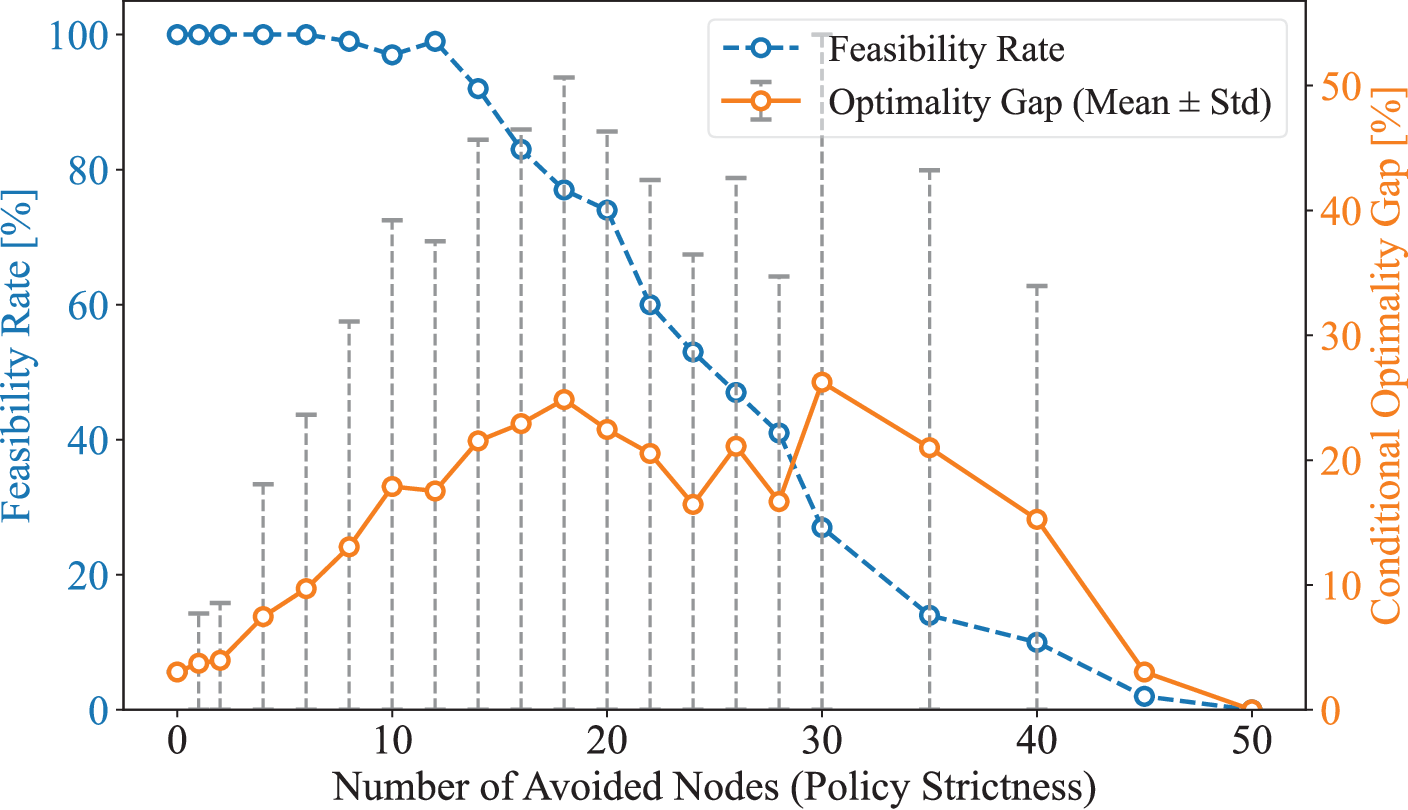}
	\caption{Evaluation of the feasibility rate and conditional optimality gap versus Operator~A's policy strictness. The number of avoided nodes $|\mathcal{V}_{a}|$ is varied from 0 to 50.}
\label{fig:result_optimal_gap_feasibility_rate}
\end{figure}

The evaluation focuses on a single time step at 2024/12/15 00:00 UTC.
The simulation settings are consistent with those in Section~\ref{section:simulation1}, except for Operator~A's policy.
In this simulation, Operator~A adopts the node-avoidance policy denoted as $\pi_{\text{AN}}(\mathcal{V}_{a})$ and we vary $|\mathcal{V}_{a}|$ across 21 discrete values ranging from 0 to 50, $\{0, 1, 2, 4, \dots, 30, 35, \dots, 50\}$, with finer granularity in the lower range (steps of 2 up to 30, and steps of 5 thereafter).
For each value, we performed 100 Monte Carlo trials.
In each trial, the set of avoided nodes $\mathcal{V}_{a}$ was randomly selected from Operator~A's 50 satellite nodes.

Fig.~\ref{fig:result_optimal_gap_feasibility_rate} illustrates these metrics as a function of the strictness of Operator~A's policy.
The horizontal axis represents the policy strictness, defined by the number of avoided nodes $|\mathcal{V}_{a}|$.
The dashed blue line (left vertical axis) represents the feasibility rate, indicating the percentage of trials where the proposed framework found at least one valid route satisfying both operators' policies.
The solid orange line (right vertical axis) represents the conditional optimality gap, plotted as the mean with error bars indicating the standard deviation.

We observe a clear trade-off between policy strictness and network performance.
The feasibility rate remains near 100\% when the number of avoided nodes is small ($|\mathcal{V}_{a}| \leq 12$).
However, as the policy becomes stricter ($|\mathcal{V}_{a}| \geq 14$), the feasibility drops sharply, approaching 0\% when $|\mathcal{V}_{a}|$ reaches 50.
This indicates that avoiding a large portion of the satellite nodes significantly reduces the probability of finding a feasible end-to-end route.
The optimality gap shows an increasing trend as policy strictness increases.
For moderate constraints ($0 \leq |\mathcal{V}_{a}| \leq 30$), the gap rises from 3.03\% to a peak of 26.24\%. 
This implies that even when a feasible route is identified, the orchestrator is forced to choose a highly sub-optimal path to bypass the blocked nodes.
The large error bars (standard deviation) indicate high variance in the optimality gap. In other words, if the blocked nodes lie on the optimal path, the delay increases significantly; otherwise, the impact is minimal.

This analysis quantitatively demonstrates the ``price of autonomy,'' where strictly imposing individual preferences degrades global system performance.
Consequently, operators should configure their policies by weighing their need for independence against global network efficiency.
This trade-off also underscores the necessity of the negotiation and relaxation mechanisms discussed in Section~\ref{section:simulation1_negotiation}, which enable the system to find a practical compromise between these conflicting objectives.

\subsection{Effects of the Number of Operators}
\label{section:simulation2}

In this subsection, the effect of the number of operators is investigated.
We set the number of operators $N_{p}$ as two, five, and ten, and the total of 100 LEO satellites is divided among the operators according to the number of operators.
The distribution of the satellites based on the number of operators is shown in Table~\ref{table:simulation_number_operators_nodes}.
The other conditions are assumed to be the same as those in Section~\ref{section:simulation1}.

\begin{table}[tb]
  \centering
  \caption{The simulation settings of the distribution of nodes and orbits based on the number of operators. There are ten LEO satellites per orbit.}
  \scalebox{1}{
    \begin{tabular}{cclc} \hline
      \multirow{2}{*}{$N_{p}$} & \# of nodes & \multirow{2}{*}{Distribution} & Difference of RAAN \\
      & per operator & & with adjacent orbit \\ \hline \hline
      2 & 50 & 5 orbits each & 72 \\
      5 & 20 & 2 orbits each & 180 \\
      10 & 10 & 1 orbit each & - \\
      \hline
    \end{tabular}
  }
  \label{table:simulation_number_operators_nodes}
\end{table}

\begin{figure}[tb]
    \centering
    \begin{minipage}[t]{0.48\hsize}
        \centering
        \subfloat[]{
            \includegraphics[width=0.93\columnwidth]{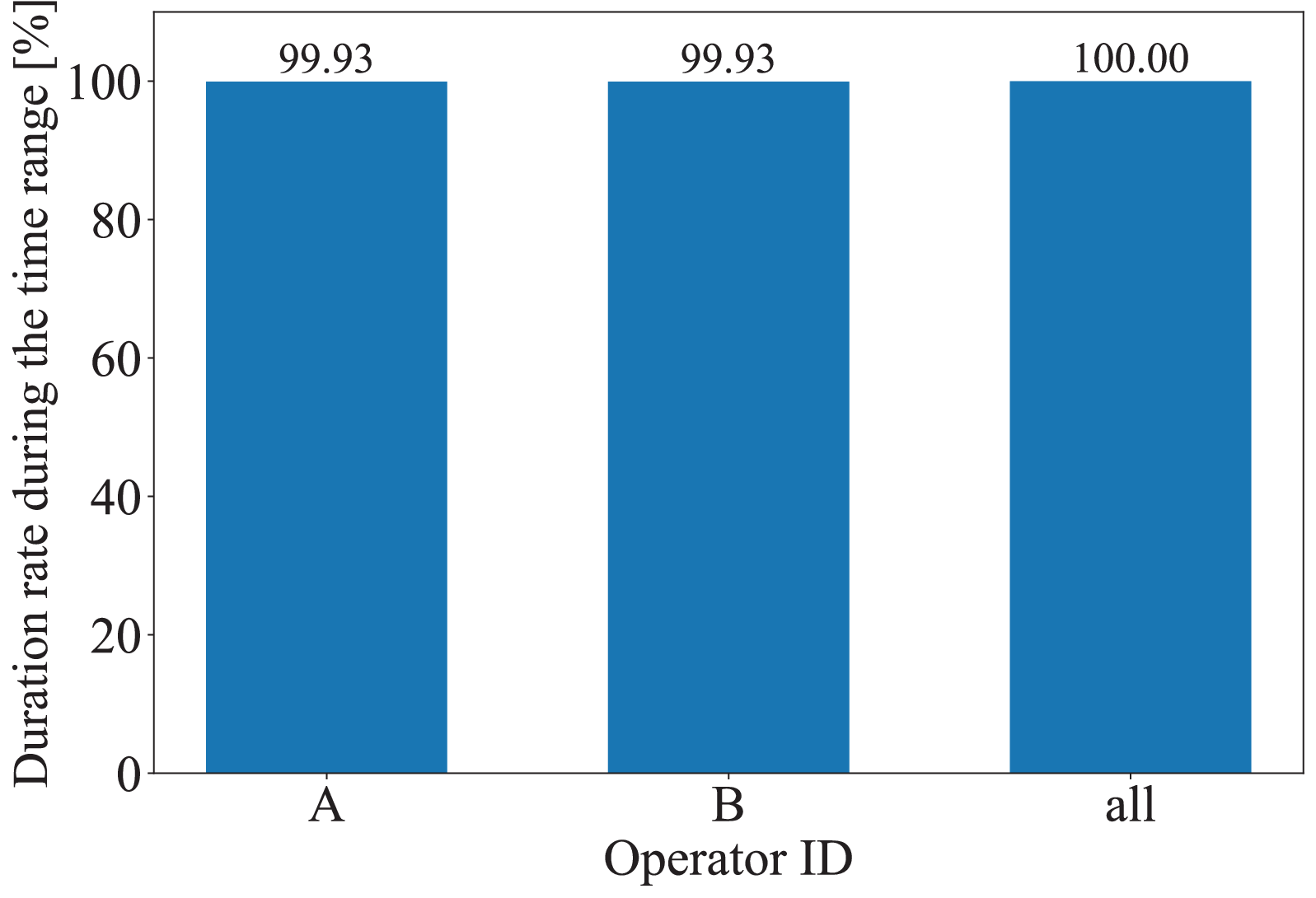}
            \label{fig:result_duration_rate_num_operators_2}
        }
    \end{minipage}
    \hfill
    \begin{minipage}[t]{0.48\hsize}
        \centering
        \subfloat[]{
            \includegraphics[width=0.93\columnwidth]{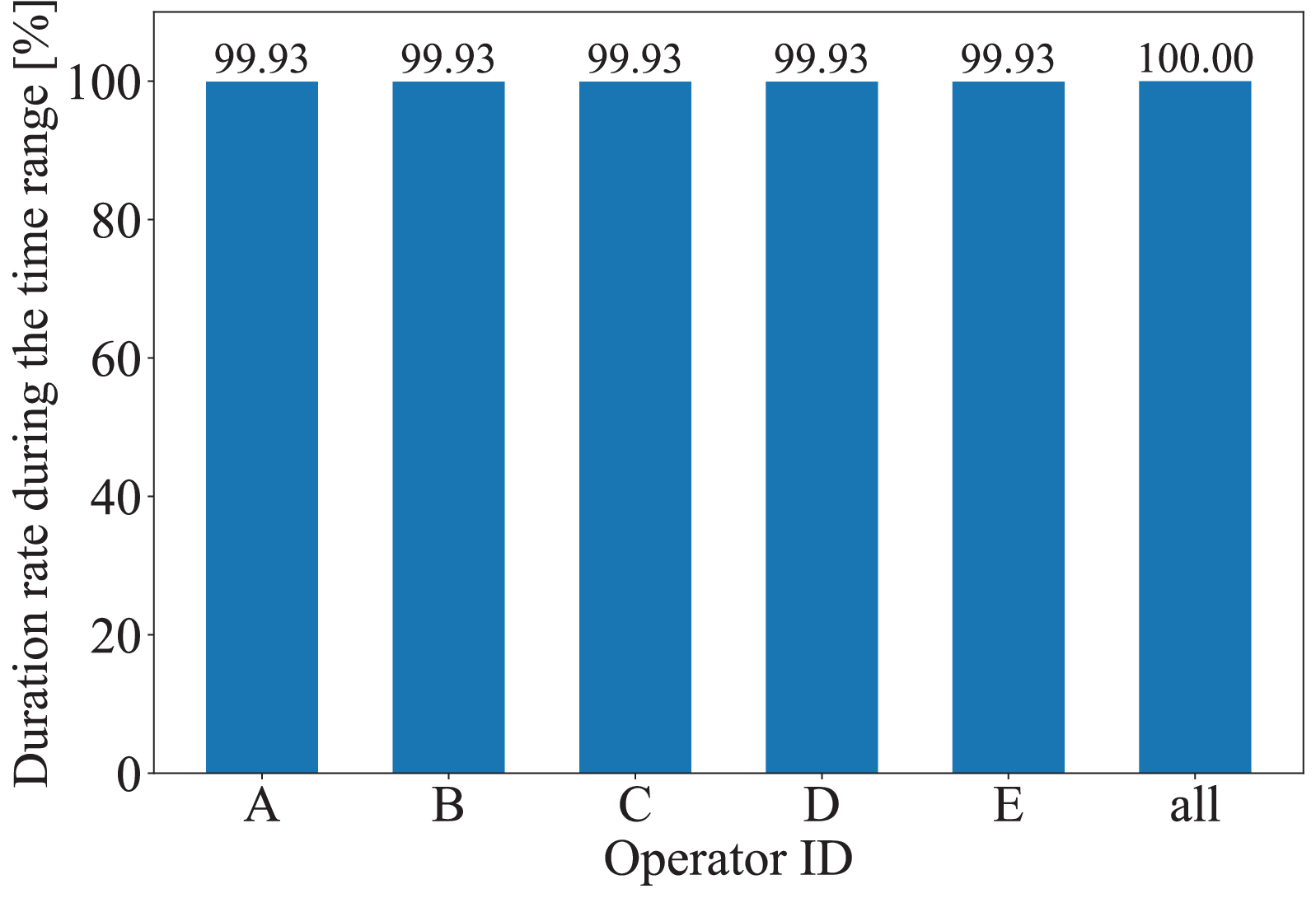}
            \label{fig:result_duration_rate_num_operators_5}
        }
    \end{minipage} \\ \smallskip
    \begin{minipage}[t]{0.8\hsize}
        \centering
        \subfloat[]{
            \includegraphics[width=\columnwidth]{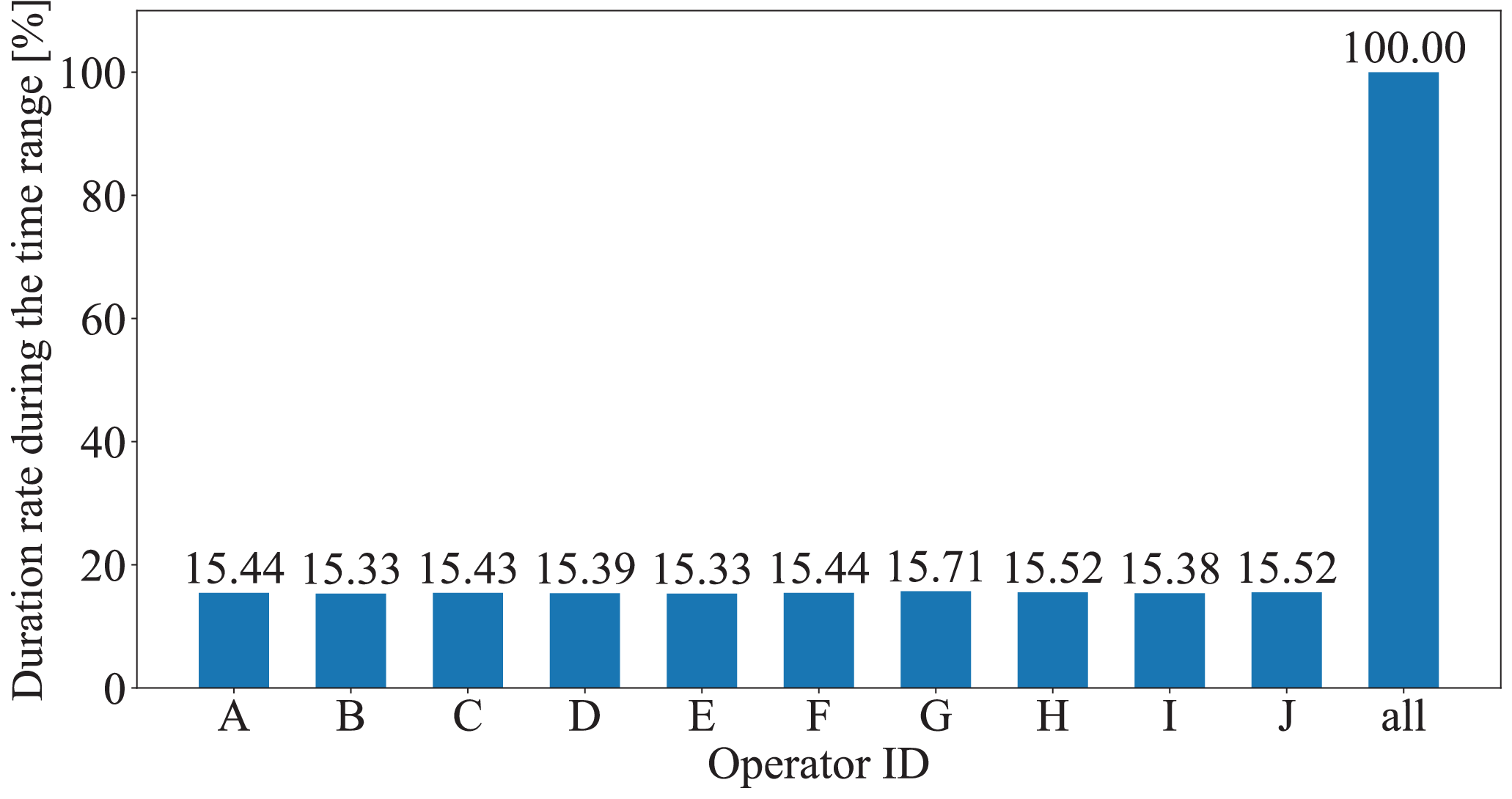}
            \label{fig:result_duration_rate_num_operators_10}
        }
    \end{minipage}
    \caption{The availability of each operator's satellite network, represented by the time ratio; (a) Two operators case, (b) Five operators case, and (c) Ten operators case.}
    \label{fig:results}
\end{figure}

\begin{table*}[tb]
  \centering
  \caption{The relationship between the number of candidate routes presented by the orchestrator, the number of routes selected by the operators, and the result of the obtained route in the scenario in Section~\ref{section:simulation2}.}
  \scalebox{0.9}{
    \begin{tabular}{ccccccccccccccc} \hline 
      \multirow{2}{*}{$N_{p}$} & \multirow{2}{*}{$N_{r}$} & \multicolumn{10}{c}{$|\bm{\Scal}_{i}^{*}|/|\bm{\Scal}_{i}^{+}|$} &  \multirow{2}{*}{$|\bm{\Rcal}^{*}|$} & \multirow{2}{*}{$n^{*}_{h}$} & \multirow{2}{*}{$\ell^{*}$} \\
      & & A ($\pi_{1}$) & B ($\pi_{2}$) & C ($\pi_{1}$) & D ($\pi_{2}$) & E ($\pi_{1}$) & F ($\pi_{2}$) & G ($\pi_{1}$) & H ($\pi_{2}$) & I ($\pi_{1}$) & J ($\pi_{2}$) & & \\ \hline \hline
      2 & 1325 & 469/1325 & 30/1325 & - & - & - & - & - & - & - & - & 5 & 5 & 46.92~ms \\
      5 & 1325 & 388/1123 & 317/1098 & 503/1039 & 239/1015 & 476/974 & - & - & - & - & - & 21 & 6 & 57.71~ms \\
      10 & 1325 & 506/815 & 152/800 & 539/862 & 129/834 & 581/836 & 212/932 & 441/928 & 129/812 & 491/731 & 147/758 & 4 & 7 & 76.56~ms\\
    \hline
    \end{tabular}
  }
  \label{table:simulation_results_num_operators_num_routes}
\end{table*}

First, we calculated the availability of each operator's satellite network, specifically the time ratio in which a route from the source to the destination could be constructed using only their own satellites.
Here, the time was set from 2024/12/15 00:00 UTC to 2024/12/16 00:00 UTC.
The results for cases where the number of operators is two, five, and ten are shown in Figs.~\ref{fig:result_duration_rate_num_operators_2}, \ref{fig:result_duration_rate_num_operators_5}, and \ref{fig:result_duration_rate_num_operators_10}, respectively.
The horizontal axis label ``all'' represents the case where all 100 satellites are used in the calculation, indicating the availability achieved through the operator collaboration we focus on in this paper.
When the number of operators is two or five, the availability of a single operator is nearly 100\%.
However, this reflects availability under the condition that at least one route exists at each time step.
Thus, operator collaboration remains beneficial if there are cases where a route cannot be established due to satellite failures or other unforeseen factors.
On the other hand, when the number of operators is ten, i.e., each operator owns only one orbital plane, the availability of an individual operator drops to approximately 15\%.
In this case, operator collaboration can increase availability to 100\%, indicating that the fewer satellites each operator owns, the more significant the benefits of collaboration become.

Next, we evaluate the effect of the number of operators in the whole network of the proposed framework.
Table~\ref{table:simulation_results_num_operators_num_routes} shows the relationship between the number of candidate routes presented by the orchestrator, the number of routes selected by the operator, and the obtained route.
Operators are labeled A and B for two operators, A to E for five operators, and A to J for ten operators.
In this simulation, we adopt two different policies for each operator; one is to avoid using the three most frequently appearing satellites included in the route candidates presented by the orchestrator for each operator, and the other is to $\pi_{\text{LH}}(10)$.
These policies are denoted by $\pi_{1}$ and $\pi_{2}$, respectively, and alternately adopted to each operator for each policy.
Here, to evaluate the number of route candidates provided to each operator, we define a subset of $\bm{\Scal}{i}$ consisting only of non-empty elements, denoted as
\begin{align}
    \bm{\Scal}_{i}^{+} := \{ \bar{\Qcal}_{i,\ell} \mid \bar{\Qcal}_{i,\ell} \neq \emptyset \} \subseteq \bm{\Scal}_{i}. \nonumber
\end{align}
This subset $\bm{\Scal}_{i}^{+}$ includes only those elements $\bar{\Qcal}_{i,\ell}$ is not empty.
Therefore, in Table~\ref{table:simulation_results_num_operators_num_routes}, the denominator and numerator of the value $|\bm{\Scal}_{i}^{*}| / |\bm{\Scal}_{i}^{+}|$ represent the number of candidate routes presented by the orchestrator that pass through each operator's network, and the number of desirable routes selected by each operator, respectively.

We compare the acceptable route rate: $|\bm{\Scal}_{i}^{*}| / |\bm{\Scal}_{i}^{+}|$.
Depending on the policy, the average acceptable route rate for the operators with $\pi_{1}$ is 0.354 for $N_{p}=2$, 0.439 for $N_{p}=5$, and 0.618 for $N_{p}=10$, and increases monotonically with the number of operators.
On the other hand, for the operators with $\pi_{2}$, the values are 0.023 for $N_{p}=2$, 0.262 for $N_{p}=5$, and 0.185 for $N_{p}=10$, indicating an increasing trend, although not a monotonic increase.
In addition, the hop count and total latency tend to increase as the number of operators increases.
This is due to the fact that an increase in the number of operators reduces the number of common routes among operators and complicates the orchestrator's route selection.

\subsection{Multi-Layered Network Case}
\label{section:simulation3}

\begin{table}[tb]
    \centering
    \caption{Simulation settings of the orchestrator and operators, their nodes, and policies in the multi-layered scenario.}
    \scalebox{1}{
    \begin{tabular}{llcl} \hline 
        Parameter & Value \\ \hline \hline
        \textbf{Orchestrator} \\
        \hline
        \multirow{2}{*}{Policy} & $\pi_{0}=\pi_{\text{LH}}(10)$ in Step~1 \\
        & $\pi_{0}=\pi_{\text{ML}}$ in Step~3 \\
        \hline
        \textbf{Operator A} \\
        \hline
        NTT node & LEO-A-1 - LEO-A-39 \\
        \# of NTN nodes & 39 \\
        Policy & $\pi_{\text{A}}=\pi_{\text{MH}}$ \\
        \hline
        \textbf{Operator B} \\
        \hline
        \multirow{2}{*}{NTN node} & LEO-B-1 - LEO-B-39 \\
        & GEO-B-1, GEO-B-2, GEO-B-3 \\
        \# of NTN nodes & 42 \\
        Policy & $\pi_{\text{B}}=\pi_{\text{MO}}$ \\
    \hline
    \end{tabular}
    }
\label{table:simulation_multilayer}
\end{table}

\begin{table}[tb]
    \centering
    \caption{Satellite orbit parameters in the multi-layered scenario.}
    \scalebox{1}{
        \begin{tabular}{ll} \hline
            Parameter & Value \\ \hline \hline
            \textbf{LEO satellite} \\ \hline
            \# of satellites in total & 78 \\
            ~~\# of planes & 6 \\
            ~~\# of satellites in each plane & 13 \\
            Altitude & 1,000~km \\
            Inclination & 55~deg \\
            Eccentricity & 0~deg \\ 
            Difference of RAAN with adjacent orbit & 60~deg \\ 
            \hline
            \textbf{GEO satellite} \\ \hline
            \# of satellites in total & 3 \\
            Longitude & 135, -140, 15~deg \\
            \hline
        \end{tabular}
    }
\label{table:simulation_parameters_multilayer}
\end{table}

\begin{table*}[tb]
  \centering
  \caption{The relationship between the number of candidate routes presented by the orchestrator, the number of routes selected by the operators, and the result of the obtained route in Section~\ref{section:simulation3}. In this table, ``\text{LEO-A-*}'' is denoted by ``\text{A-*}'' for simplicity.}
  \scalebox{0.86}{
    \begin{tabular}{cccccccclcc} \hline 
        ID & $\pi_{0}$ (Step~1) & $\pi_{\text{A}}$ & $\pi_{\text{B}}$ & $N_{r}$ & $|\bm{\Scal}^{*}_{\text{A}}|$ & $|\bm{\Scal}^{*}_{\text{B}}|$ & $|\bm{\Rcal}^{*}|$ & Resulting route & $n^{*}_{h}$ & $\ell^{*}$ \\ \hline \hline

    \hline
    \multicolumn{11}{l}{\textbf{Centralized Framework}} \\
    \hline
      
      C-1 & - & - & - & - & - & - & - & 
      $\text{User} \rightarrow \underbrace{\text{A-32}}_{\text{Operator~A}} \rightarrow \underbrace{\text{B-19}}_{\text{Operator~B}} \rightarrow \underbrace{\text{A-19}}_{\text{Operator~A}} \rightarrow \underbrace{\text{B-6}}_{\text{Operator~B}} \rightarrow \text{OGS} \rightarrow \text{DN}$ & 
      6 & 48.86~ms \\

    \hline
    \multicolumn{11}{l}{\textbf{Proposed Framework}} \\
    \hline
      
      P-1 & $\pi_{\text{LH}}(10)$ & $\pi_{\text{MH}}$ & $\pi_{\text{MO}}$ & 1757 & 79 & 243 & 37 & 
      $\text{User} \rightarrow \underbrace{\text{B-31} \rightarrow \text{B-30} \rightarrow \text{B-29}}_{\text{Operator~B}} \rightarrow \underbrace{\text{A-6}}_{\text{Operator~A}} \rightarrow \text{OGS} \rightarrow \text{DN}$ & 
      6 & 53.80~ms \\
      
      P-2 & $\pi_{\text{LH}}(10) \land \pi_{\text{LL}}(120~\text{ms})$ & $\pi_{\text{MH}}$ & $\pi_{\text{MO}}$ & 1459 & 51 & 200 & 32 & 
      $\text{User} \rightarrow \underbrace{\text{B-31} \rightarrow \text{B-30} \rightarrow \text{B-29}}_{\text{Operator~B}} \rightarrow \underbrace{\text{A-6}}_{\text{Operator~A}} \rightarrow \text{OGS} \rightarrow \text{DN}$ & 
      6 & 53.80~ms \\
      
      \hline
    \end{tabular}
  }
  \label{table:simulation_results_multilayer}
\end{table*}

In this scenario, we introduce three GEO satellites into the network to model a multi-layered network and evaluate the performance of the proposed framework.
Tables~\ref{table:simulation_multilayer} and \ref{table:simulation_parameters_multilayer} present the simulation settings for this subsection.
We use all the link parameters of LEO satellites, GEO satellites, and OGS listed in Table~\ref{table:simulation_link_parameters} in Appendix~\ref{section:link_parameters}.
We assume the presence of an orchestrator and two NTN operators ($N_{p}=2$), where one operator owns only LEO satellites, and the other also owns GEO satellites.
The total number of LEO satellites is set to 78, distributed across six orbital planes with 13 satellites per plane.
The GEO satellites are positioned at longitudes of 135, -140, and 15~deg.

First, as in Section~\ref{section:simulation1}, we show the result of the centralized orchestration framework.
The result of the route obtained by the centralized orchestration framework is shown in C-1 of Table~\ref{table:simulation_results_multilayer}, where the hop count was six and the latency was 48.86~ms.
Note that there are three inter-operator connections: two from Operator~A to B, and one from Operator~B to A.

In the proposed orchestration framework, the orchestrator first calculates candidate routes with the policy of $\pi_{0}=\pi_{\text{LH}}(10)$, and the operators set their policies as $\pi_{\text{A}}=\pi_{\text{MH}}$ and $\pi_{\text{B}}=\pi_{\text{MO}}$, and the latter meaning that fewer inter-operator connections are preferable.
The result was shown in P-1 of Table~\ref{table:simulation_results_multilayer}, where the hop count was six and the latency was 53.80~ms.
The number of the candidate routes and the number of routes satisfying the requirements of Operators A, B, and both of two operators were $N_{r}=1757$, $|\bm{\Scal}^{*}_{\text{A}}|=79$, $|\bm{\Scal}^{*}_{\text{B}}|=243$, and $|\bm{\Rcal}^{*}|=37$, respectively.

\begin{figure}[tb]
	\centering
	\includegraphics[width=0.95\columnwidth]{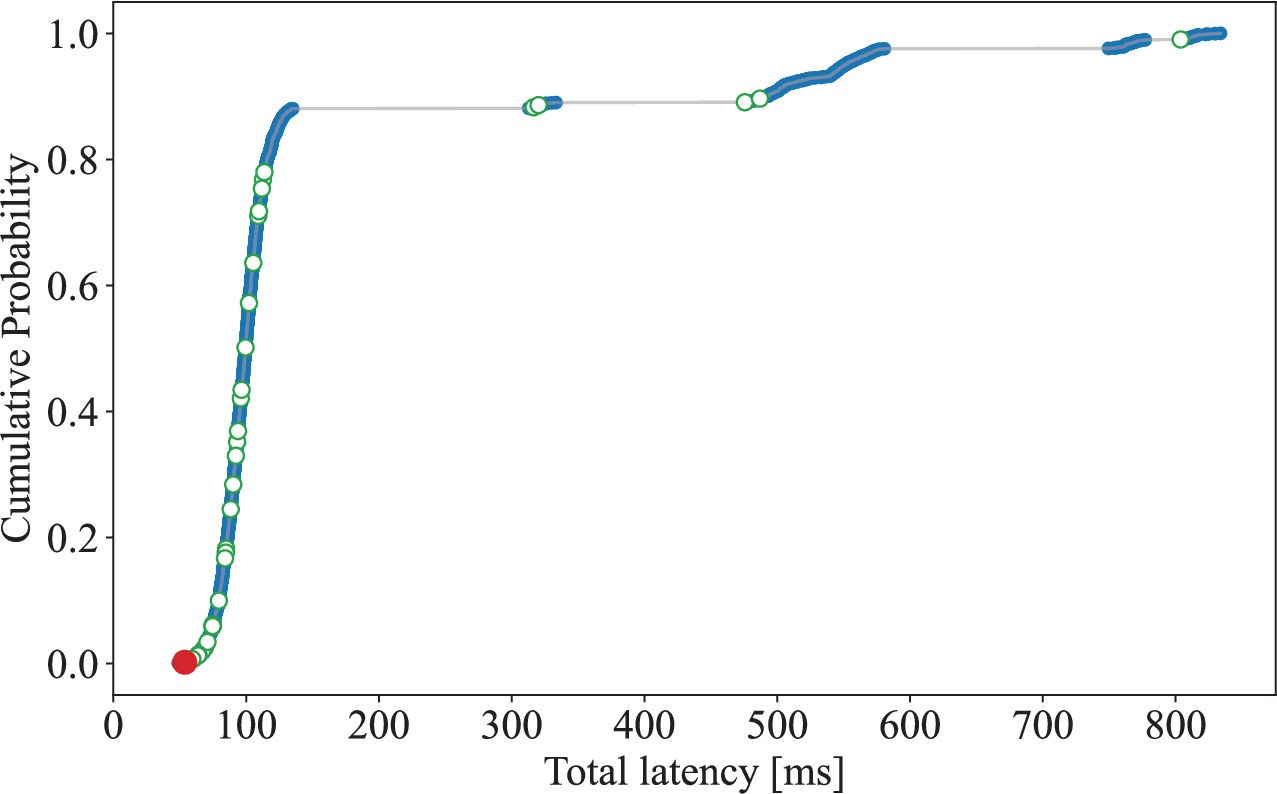}
	\caption{Cumulative distribution function (CDF) of the total latency for candidate routes presented by the orchestrator with $\pi_{\text{LH}}(10)$. Blue, green, and red circles represent the route candidates provided by the orchestrator, the intersection of the selected routes from both operators among the candidates, and the route determined by the proposed framework, respectively.}
\label{fig:result_latency_multilayer}
\end{figure}

The key characteristic of this scenario is that it includes GEO satellites, which have longer latency compared to LEO satellites.
Fig.~\ref{fig:result_latency_multilayer} shows the cumulative distribution function (CDF) of the total latency for candidate routes presented by the orchestrator.
The horizontal axis represents the total latency, and the vertical axis represents the cumulative probability.
Blue, green, and red circles represent the route candidates provided by the orchestrator, the intersection of the selected routes from both operators among the candidates, and the route determined by the proposed framework, respectively.
Significant gaps in latency are observed in three ranges (135.10--312.80~ms, 333.37--475.63~ms, and 580.65--749.48~ms), indicating that these routes pass through one, two, and three GEO satellites, respectively.
As we have explained, the proposed orchestration framework allows for trade-offs between the orchestrator's and individual operators' policies.
However, routes that include the GEO satellite are selected, leading to longer latency compared to that of the centralized orchestration framework.
In this scenario, the total latency can range from 300 to 800~ms, because some green circles have a latency exceeding 300~ms.
Therefore, the orchestrator may then change the policy with a latency threshold and present each operator with only routes that satisfy that condition.

\begin{figure}[tb]
	\centering
	\includegraphics[width=0.95\columnwidth]{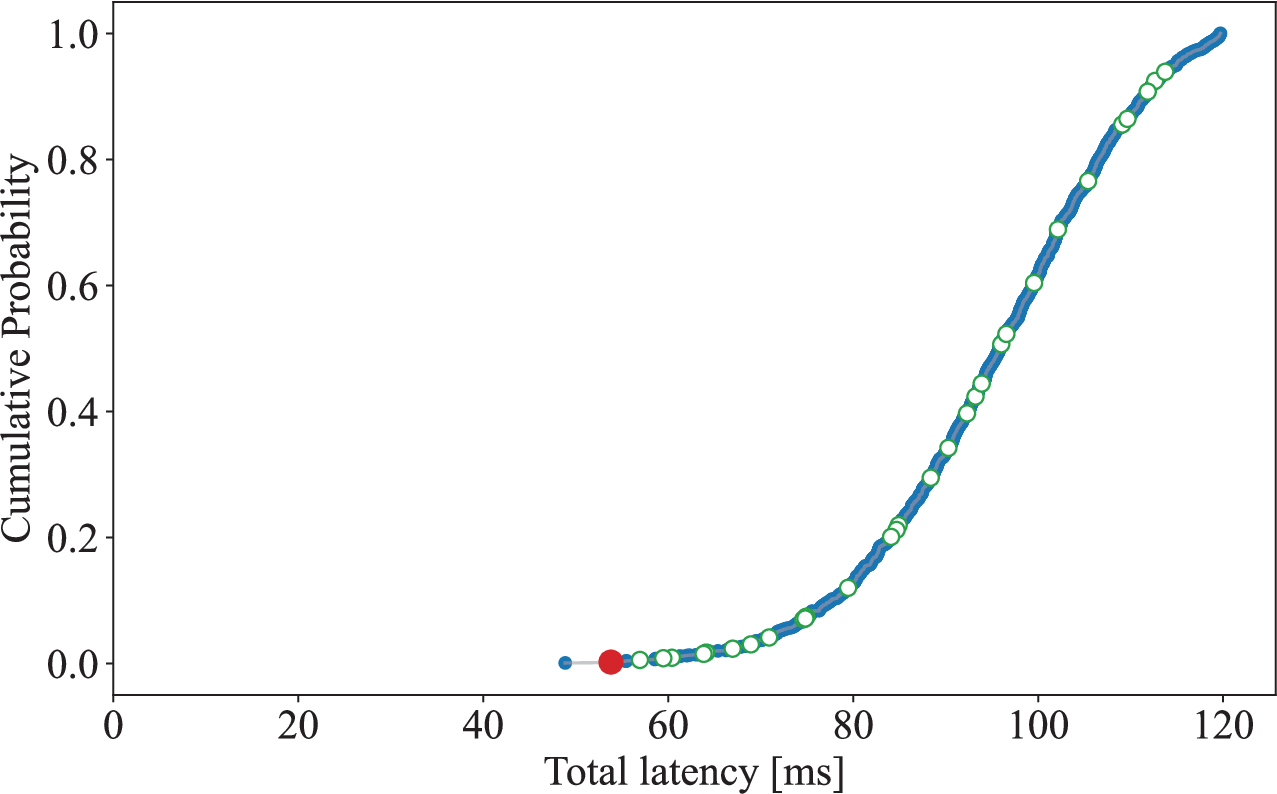}
	\caption{CDF of the total latency for candidate routes presented by the orchestrator with $\pi_{0}=\pi_{\text{LH}}(10) \land \pi_{\text{LL}}(120~\text{ms})$. The legend for the plots is the same as in Fig~\ref{fig:result_latency_multilayer}.}
\label{fig:result_latency_multilayer_latency_constraint}
\end{figure}

We examine it using this scenario as an example.
Here, we set a policy of the orchestrator not only $\pi_{\text{LH}}(10)$ but also $\pi_{\text{LL}}(120~\text{ms})$.
Therefore, the policy of the orchestrator in Step~1 is ``Select routes with the hop count less than or equal to ten and latency less than or equal to 120~ms.''
The result obtained by the proposed framework was written in P-2 of Table~\ref{table:simulation_results_multilayer}.
The resulting route is the same as the previous result, and the number of candidate routes decreased to 1459.
Fig.~\ref{fig:result_latency_multilayer_latency_constraint} shows the CDF of the latency for the candidate routes after the orchestrator adds the policy $\pi_{\text{LL}}(120~\text{ms})$.
From this figure, we can observe that the latency of the candidate routes presented by the orchestrator does not exceed 120 ms.
Finally, we evaluate the latency of the candidate routes that could have been selected as the best route, the routes marked with green circles in these figures.
Under the conditions before introducing the additional policy, the maximum latency of these routes was 803.82~ms (see Fig.~\ref{fig:result_latency_multilayer}).
However, after introducing the additional policy, the maximum latency was reduced to 113.74~ms (see Fig.~\ref{fig:result_latency_multilayer_latency_constraint}).
Compared to the centralized framework, which achieved a latency of 48.86~ms, the latency increase was significantly reduced from 754.96~ms to 64.88~ms.
Therefore, by adding the latency constraint for the orchestrator's policy, the proposed method also allows for operations that actively avoid long-latency routes that are undesirable for users.

\section{Conclusion}
\label{section:conclusion}

In this paper, we proposed a network orchestration framework to connect multiple NTN operators with only a few NTN nodes and to construct a virtual constellation.
In the proposed framework, the orchestrator calculates candidate routes based on the node and link information provided by each operator and notifies the operators of these options.
Each operator then selects preferred routes from the candidates based on its own policies and responds to the orchestrator.
The orchestrator determines the final route through optimization.
Unlike a centralized orchestration framework, this approach enables route design in NTN that takes into account each operator's operational policies.

We confirmed the effectiveness of the proposed framework through the numerical simulations.
Specifically, we evaluated five scenarios: (i) a case with two operators, each owning an LEO constellation; (ii) a case where the number of operators increased to two, five, and ten; and (iii) a case where GEO satellites were also present in the network.
Furthermore, using the network configuration from scenario (i), we extended the evaluation to a dynamic NTN environment to verify the framework's adaptability to time-varying topologies.
In this context, we also analyzed the iterative negotiation process to address policy conflicts and quantitatively demonstrated the ``price of autonomy,'' highlighting the trade-off between strict operator policies and global system performance.
These findings highlighted the practical advantages of orchestration in building collaborative networks among multiple operators in NTN and demonstrated its potential to contribute to the development of future NTN environments.

Looking forward, the proposed orchestration framework aligns well with emerging architectural paradigms for next-generation NTNs.
Specifically, the orchestrator's capability to coordinate the virtual global network exhibits conceptual synergy with inter-operator network slicing~\cite{Nguyen_OJCS24} and digital twin frameworks~\cite{Zhou_CM24}, while the distributed nature of the operators supports edge computing~\cite{Zhou_TMC24} and privacy-preserving techniques such as federated learning~\cite{Shi_TWC24}.
Integrating these technologies into our framework is a promising direction for future work.

\section*{Acknowledgment}

This study is conducted under the commissioned research of the ``Research and Development of Ka-band Satellite Communication Control for Various Use Cases'' (JPJ000254) by the Ministry of Internal Affairs and Communications, Japan.

\appendices

\section{Link Parameters Used in the Simulations}
\label{section:link_parameters}

This appendix summarizes the link parameters used in the simulations in Section~\ref{section:simulation}.
Sections~\ref{section:simulation1} and \ref{section:simulation2} use LEO satellite and OGS parameters, whereas Section~\ref{section:simulation3} uses all parameters listed in Table~\ref{table:simulation_link_parameters}.
Optical links were assumed for inter-satellite and satellite-to-OGS communications, with received power calculated using Eq.~(\ref{eq:received_power}) in Section~\ref{section:system_model}.
For user-to-satellite communications, RF links were assumed; however, in this paper, the user was connected to the satellite with the shortest distance among the candidates.

\begin{table}[tb]
    \centering
    \caption{Link parameters used in the simulations.}
    \scalebox{1}{
        \begin{tabular}{ll} \hline
            Parameter & Value \\ \hline \hline
            \textbf{Common parameters} \\ \hline
            Wavelength $\lambda$ (for optical links) & 1550~nm \\
            Other losses $L_{s}$ & 0~dB \\
            \hline
            \textbf{Feasible link condition} \\ \hline
            Distance threshold $D_{\text{req}}$ & 10,000~km \\
            Required received power $P_{r, \text{req}}$ & -50~dBm \\
            \hline
            \textbf{LEO satellite} \\ \hline
            Transmit power $P_{t}$ & 30~dBm \\
            Transmit antenna gain $G_{t}$ & 106~dBi \\
            Receive antenna gain $G_{r}$ & 106~dBi \\
            \hline
            \textbf{GEO satellite} \\ \hline
            Transmit power $P_{t}$ & 37~dBm \\
            Transmit antenna gain $G_{t}$ & 114~dBi \\
            Receive antenna gain $G_{r}$ & 114~dBi\\
            \hline
            \textbf{OGS} \\ \hline
            Receive antenna gain $G_{r}$ & 118~dBi \\
            \hline
        \end{tabular}
    }
\label{table:simulation_link_parameters}
\end{table}

\bibliography{myref.bib}

\end{document}